\documentclass{jfm}
\usepackage[utf8]{inputenc}
\usepackage{natbib}
\usepackage{amssymb,amsmath}
\usepackage{epsfig}
\usepackage{bm}
\usepackage{color}

\usepackage{graphicx}

\usepackage{color}
\newcommand{\vicente}[1]{{  #1}}

\newcommand\beq{\begin{equation}}
\newcommand\eeq{\end{equation}}
\newcommand\beqa{\begin{eqnarray}}
\newcommand\eeqa{\end{eqnarray}}

\newcommand{\al}{\alpha}

\title[Diffusion of intruders in a granular suspension]{Diffusion of intruders in a granular gas thermostatted by a bath of elastic hard spheres}

\author[Rub\'en G\'omez Gonz\'alez and Vicente Garz\'o]%
{R\ls U\ls B\ls \'E\ls N\ls \ns G\ls \'O\ls M\ls E\ls Z G\ls O\ls N\ls Z\ls \'A\ls L\ls E\ls Z\ls$^1$ and
V\ls I\ls C\ls E\ls N\ls T\ls E\ls \ns G\ls A\ls R\ls Z\ls \'O\ls$^2$}

\affiliation{$^1$Departamento de
Did\'actica de las Ciencias Experimentales y las Matem\'aticas, Universidad de Extremadura, E-10004 C\'aceres, Spain
\\[\affilskip]
$^2$ Departamento de F\'isica, Instituto Universitario de Computaci\'on Cient\'ifica Avanzada (ICCAEx), Universidad de Extremadura, Avda. de Elvas s/n, 06006  Badajoz (Spain).}

\begin{document}

\maketitle

\begin{abstract}

The Boltzmann kinetic equation is considered to compute the transport coefficients associated with the mass flux of intruders in a granular gas. Intruders and granular gas are immersed in a gas of elastic hard spheres (molecular gas). We assume that the granular particles are sufficiently rarefied so that the state of the molecular gas is not affected by the presence of the granular gas. Thus, the gas of elastic hard spheres can be considered as a thermostat (or bath) at a fixed temperature $T_g$. In the absence of spatial gradients, the system achieves a steady state where the temperature of the granular gas $T$ differs from that of the intruders $T_0$ (energy nonequipartition). Approximate theoretical predictions for the temperature ratio $T_0/T_g$ and the kurtosis $c_0$ associated with the intruders compare very well with Monte Carlo simulations for conditions of practical interest. For states close to the steady homogeneous state, the Boltzmann equation for the intruders is solved by means of the Chapman--Enskog method to first order in the spatial gradients. As expected, the diffusion transport coefficients are given in terms of the solutions of a set of coupled linear integral equations which are approximately solved by considering the first-Sonine approximation. In dimensionless form, the transport coefficients are nonlinear functions of the mass and diameter ratios, the coefficients of restitution, and the (reduced) bath temperature. Interestingly, previous results derived from a suspension model based on an effective fluid-solid interaction force are recovered when $m/m_g\to \infty$ and $m_0/m_g\to \infty$, where $m$, $m_0$, and $m_g$ are the masses of the granular, intruders, and molecular gas particle, respectively. Finally, as an application of our results, thermal diffusion segregation is exhaustively analysed.
\end{abstract}

\date{\today}
\maketitle

\section{Introduction}
\label{sec1}

One of the most relevant characteristics of granular systems is that they are constituted by macroscopic particles (or grains) that collide inelastically among themselves. Due to this fact, the kinetic energy of the system decreases over time. Thus, to observe sustained diffusive motion of the grains, an external energy input is usually introduced to compensate for the energy lost by collisions and reach a nonequilibrium steady state. Several mechanisms are used to inject energy at the system in the real experiments, e.g., mechanical-boundary shaking \cite[]{YHCMW02,HYCMW04}, bulk driving [as in air-fluidized beds \cite[]{SGS05,AD06}], or magnetic forces \cite[]{SHKZP13,HTWS18}. However, since in most of the experimental realizations the formation of large spatial gradients in the bulk region goes beyond the Navier--Stokes domain, it is quite difficult to provide a rigorous theoretical treatment of these sort of situations. In computer simulations, the above obstacle can be circumvented by the introduction of external forces (or thermostats) \cite[]{EM90} that heat the system and compensate for the energy dissipated by collisions. Unfortunately, it is not clear so far the relation between each specific type of thermostat with experiments.

A more realistic example of thermostated granular systems consists of a set of solid particles immersed in an interstitial fluid of molecular particles. This provides a suitable starting point to mimic the behaviour of real suspensions. Needless to say, the understanding of the flow of solid particles in one or more fluid phases is in fact a quite intricate problem. Among the different types of multiphase flows, a simple but interesting set corresponds to the so-called particle laden-suspensions \cite[]{S20}. In this sort of suspension, a set of small and dilute particles are immersed in a carrier fluid (such as water or air). When the dynamics of grains in gas-solid flows are essentially dominated by collisions, the extension of the conventional kinetic theory of gases \cite[]{CC70,FK72} to dissipative dynamics can be considered as a reliable tool to describe this sort of systems. However, at a kinetic level, the description of flows involving two or more phases is really a complex problem since one should start from a set of kinetic equations for each one of the velocity distribution functions of the different phases. In addition, the different phases evolve over quite different spatial and temporal scales. Due to these difficulties, a coarse-grained approach is usually adopted and the influence of gas-phase effects on the dynamics of solid particles is incorporated in the starting kinetic equation in an effective way by means of a fluid-solid interaction force \cite[]{K90,G94,J00}. In some cases, a Stokes linear drag law for gas-solid interactions is only accounted for \cite[]{LMJ91,TK95,SMTK96,WZLH09,H13,WGZS14,ChVG15,SA17,ASG19,SA20,ChBCh23}. Other models include an additional Langevin stochastic term
\cite[]{GTSH12,HTG17,GGG19a,GKG20,G23}.

Although the effective suspension models based on the Langevin-like equation provides a reliable way of capturing the impact of gas-phase on the dynamic properties of grains, it could be convenient from a more fundamental point of view to begin with a model that accounts for the effect of the (real) collisions between solid and gas particles. In this context, inspired in a paper reported by \cite{BMP02a} a recent (discrete) suspension model has been recently proposed \cite[]{GG22}. As in the case of the most effective models reported in the granular literature \cite[]{LMJ91,TK95,SMTK96,WZLH09,GTSH12,H13,WGZS14,ChVG15,HTG17,SA17,ASG19,SA20,GGG19a,GKG20}, the model is based on the following assumptions. First, one assumes that the granular particles are sufficiently dilute so that the state of the interstitial gas is not affected by the presence of the grains. This means that the molecular surrounding gas can be treated as a bath (or thermostat) of elastic hard spheres at a constant temperature $T_g$. This assumption can be clearly justified in the case of particle-laden suspensions where the granular particles are sufficiently rarefied. Second, although the density of solid particles is very small, the grain-grain collisions (which are inelastic and characterized by the coefficient of restitution $\al$) are accounted for in the kinetic equation for the velocity distribution function $f(\mathbf{r}, \mathbf{v}; t)$ of grains. Thus, it is quite obvious that this suspension model (granular particles immersed in a molecular gas of elastic hard spheres) can be seen as a binary mixture where one of the species (the grains) is present in tracer concentration. In the homogeneous state, a steady state is reached when the energy lost by grains (due to their inelastic collisions) is exactly compensated for by the energy gained by them due to their elastic collisions with gas particles \cite[]{BMP02a,S03a}.

It is worth noting that the suspension model introduced by \cite{GG22} has some features in common with the microscopic theory of transport for dilute molecular suspensions, as reported by \cite{SS84a,SS84b} years ago. In this theory, the dynamics of the solute-solvent collision is treated within the Enskog approximation, conveniently modified by the presence of the solvent sea. The solvent is treated as a continuum using appropriate generalised boundary conditions. These conditions allow the diffusion coefficient to properly account for dynamic memory (repeated collision events), which is neglected in the conventional Enskog theory. Additionally, the solute particles are sufficiently dilute so that the interaction between them may be neglected, yet concentrated enough to permit a statistical treatment. Their theoretical expression for the self-diffusion coefficient is in excellent agreement with molecular dynamics (MD) simulations \cite[]{AGW70}. However, the microscopic theory of \cite{SS84a,SS84b} differs from the suspension model employed by \cite{GG22}. Firstly, the theory of \cite{SS84a,SS84b} is for elastic collisions, whereas the model of \cite{GG22} considers the effect of the \textit{inelastic} collisions between the solute (grains) particles on its distribution function. Secondly, the theory of \cite{SS84a,SS84b} considers finite values of the solid volume fraction of the solvent, whereas the suspension model introduced by \cite[]{GG22} is restricted to the low-density regime. In this density regime, the inelastic Boltzmann kinetic equation applies, and it is justified to neglect the effect of dynamic correlations in repeated collisions on the transport coefficients. In this context, it is important to recall that for moderate densities the corresponding version of the inelastic Enskog equation (which goes beyond the Boltzmann description) can still be considered as a good approximation for obtaining the transport coefficients of dense granular fluids since the Enskog results \cite[]{GD99a,GD99b,GDH07,GHD07} have been shown to compare quite well with MD simulations \cite[]{LBD02,DHGD02,LLC07,MDCPH11,ChS13,MGH14} and with experimental data \cite[]{YHCMW02,HYCMW04} for moderately high densities and values of $\alpha\gtrsim 0.8$.

In contrast to coarse-grained approaches for granular suspensions, the model proposed by \cite{GG22} introduces two new input parameters: the diameter $\sigma/\sigma_g$ and mass $m/m_g$ ratios. Here, $\sigma_g$ and $m_g$ are the diameter and mass of the particles of the surrounding molecular gas, respectively, while $\sigma$ and $m$ are the diameter and mass of the solid particles, respectively. For small spatial gradients, this suspension model has been solved by means of the Chapman--Enskog method \cite[]{CC70} and the expressions of the Navier--Stokes transport coefficients of the granular suspension have been explicitly obtained in terms of the parameter space of the system \cite[]{GG22}. An interesting result is that the Navier--Stokes expressions derived from this collisional model reduce to those previously derived from a coarse-grained approach \cite[]{GGG19a} when the particles of the molecular gas are much lighter than the granular particles (Brownian limit, $m_g/m\to 0$). This agreement may justify the use of
this sort of effective Langevin-like models for obtaining the dynamic properties of grains when $m \gg m_g$ \cite[]{PZ23}.

While the study of transport properties in granular suspensions reported in \cite{GG22} has been restricted to a monocomponent granular suspension (granular gas immersed in a molecular gas), extending this analysis to the more realistic case of a bidisperse granular suspension  presents significant new conceptual and technical difficulties  and is far from straightforward. Indeed, the evaluation of Navier-Stokes transport coefficients for multicomponent suspensions introduces significant new challenges. Not only does the number of relevant transport coefficients increase due to the complexity of particle interactions in mixtures, but also these coefficients are defined by a set of coupled integro-differential equations. Furthermore, new parameters emerge, such as the mass and size ratios, along with the coefficients of restitution for each pairwise collision, making the problem substantially more intricate than in the monocomponent case.  
Thus, to gain some insight into the general problem, we will make in this paper a first step in the understanding of transport in multicomponent granular suspensions: we consider a granular binary mixture (immersed in a molecular gas) where the concentration of one of species (impurities or intruders) is much smaller than the other one (tracer limit). As mentioned before, in the tracer limit one can assume that (i) the state of the excess species (granular gas) is not perturbed by the presence of intruders and (ii) one can also neglect collisions among tracer particles themselves in their corresponding kinetic equation.

At a kinetic level, the tracer limit greatly simplifies the application of the Chapman–Enskog method \cite[]{CC70} to bidisperse granular suspensions since the transport properties of the excess species (the pressure tensor and the heat flux) are the same as that for the monocomponent granular suspension. These transport coefficients were already derived by \cite{GG22}. 
Consequently, the mass transport of impurities $\mathbf{j}_0$ is the relevant flux of the problem. In accordance with the results of tracer diffusion in granular gases \cite[]{G19}, one expects that the Navier–Stokes constitutive equation for the mass flux (that is, linear in the spatial gradients) can be written as
\beq
\label{0.1}
\mathbf{j}_0^{(1)}=-\frac{m_0^2}{\rho}D_0 \nabla n_0-\frac{m m_0}{\rho}D \nabla n-\frac{\rho}{T}D_T \nabla T-D_0^U \Delta \mathbf{U},
\eeq
where 
$\rho=m n$ is the mass density of the granular gas, $n_0$ is the number density of the intruders, $n$ is the number density of the particles of the granular gas, $T$ is the granular temperature, and $\Delta \mathbf{U}=\mathbf{U}-\mathbf{U}_g$, $\mathbf{U}$ and $\mathbf{U}_g$ being the mean flow velocities of the granular and molecular gases, respectively. In addition,
$D_0$ is the kinetic (tracer) diffusion coefficient, $D$ is the mutual diffusion coefficient, $D_T$ is the thermal diffusion coefficient, and $D_0^U $ is the velocity diffusion coefficient. While the three first diffusion coefficients are the coefficients of proportionality between the mass flux and hydrodynamic gradients, the coefficient $D_0^U$ links the mass flux with the velocity difference $\Delta \mathbf{U}$. Although this latter contribution to the mass flux does not appear in dry granular mixtures, it is also present in the heat flux of a granular suspension composed by two different phases (a granular gas immersed in a molecular gas) \cite[]{GG22}. Here, as it will be shown later, by symmetry reasons the mass flux $\mathbf{j}_0^{(1)}$ is also expected to be coupled to $\Delta \mathbf{U}$.

The determination of the diffusion transport coefficients $D_0$, $D$, $D_T$, and $D_0^U$ is the main goal of the present paper. As usual for elastic \cite[]{CC70} and inelastic \cite[]{G19} collisions, these transport coefficients are given in terms of a set of coupled linear integral equations (see the supplementary material). These integral equations are approximately solved by considering the leading terms in a Sonine polynomial expansion. However, as occurs in the case of a monocomponent granular suspension \cite[]{GG22}, evaluating the diffusion coefficients for general unsteady conditions requires numerically solving a set of nonlinear differential equations. 
In the bidisperse case, these equations differ fundamentally from the monocomponent case due to the presence of two mechanically different species, resulting in the emergence of additional parameters.
To simplify the analysis and obtain analytical results, we focus here on steady-state conditions. This enables us to get analytical results and express the diffusion transport coefficients in terms of the parameter space of the system.

The above set of diffusion transport coefficients has been recently determined in two different systems. Thus, in \cite{GGGBS24} we considered a collisional model (the so-called $\Delta$-model) to analyse the density flux of tracer particles in a confined, quasi-two-dimensional, moderately dense granular gas of inelastic hard spheres. More relevant to the present work, the diffusion coefficients of a binary granular suspension where one of the species (of mass $m_0$) is present in tracer concentration have been determined by solving the set of (inelastic) Enskog equations \cite[]{GG23}. In contrast to the suspension model considered here, a coarse-grained approach was adopted, whereby the influence of the interstitial fluid on grain motion was accounted for via effective forces (Langevin-like model). This simplification allowed us to derive explicit forms for the diffusion coefficients up to the second Sonine approximation. When $m_0$ is much greater than $m_g$, the results obtained in this study (which apply for arbitrary values of the mass ratio $m_0/m_g$) reduce to those derived in \cite{GG23} in the low-density regime and when only the first Sonine approximation is considered. In this sense, the present work subsumes previous studies \cite[]{GG23} that are recovered in some limiting cases ($m_0/m_g \to \infty$ and $m/m_g \to \infty$).

Given that the explicit forms of the diffusion coefficients are at hand, as an interesting application of our results, we derive a segregation criterion for the intruders based on the knowledge of the so-called thermal diffusion factor (see, e.g., \cite{GI52,GR83a,GR83b,KCL87,BRM05,G06,G08a,BS09,GG23,GGGBS24}). Segregation is induced here by both gravity and a temperature gradient. Three different situations are considered: one without gravity, another dominated by gravity, and an intermediate case. Surprisingly, the segregation dynamics found here differ from those derived by using a Langevin-like approach in \cite{GG23}. However, despite the plots appearing so different, we can explain those differences.
They stem essentially from the way the molecular gas thermalizes the grains in our (discrete) suspension model, which contrasts with the effective thermostat used in the coarse-grained approaches \cite[]{GG23}. Additionally, our model captures the full mass ratio dependence and therefore reveals how segregation varies as a function of the mass ratios $m_0/m_g$ and $m/m_g$, offering a more general description beyond the Brownian limit ($m_0/m_g \to \infty$ and $m/m_g \to \infty$) considered in \cite{GG23}. This is in fact one of the new added values of the present work.

The structure of the paper is as follows. Section \ref{sec2} introduces the Boltzmann kinetic equation for granular particles immersed in a molecular gas and analyzes the homogeneous steady state. In Section \ref{sec3}, some intruders are added to the granular gas and the corresponding Boltzmann-Lorentz kinetic equation is derived. We first consider the homogeneous steady state for intruders and show how the non-equipartition of energy is affected by the mass ratio $m_0/m_g$. Section \ref{sec4} presents the set of integral equations governing the diffusion transport coefficients, while section \ref{sec5} provides approximate expressions (based on the so-called first Sonine approximation) for these coefficients. These coefficients are explicitly determined in terms of the background temperature, volume fraction, restitution coefficients, and the masses and diameters of the bidisperse system. Six appendices in the Supplementary Material provide technical details of the calculations and simulation techniques. The convergence to the results obtained by \cite{GG22a} from the Langevin-like model is also demonstrated. Section \ref{sec7} examines thermal diffusion segregation. The paper concludes in section \ref{sec8} with a brief discussion of the results reported in this paper.

\section{Granular gas in contact with a bath of elastic hard spheres. Boltzmann kinetic description}
\label{sec2}

We consider a gas of inelastic hard disks ($d=2$) or spheres ($d=3$) of mass $m$, diameter $\sigma$, and coefficient of normal restitution $\al$. We assume that the spheres
are perfectly smooth and therefore the collisions between particles are inelastic and characterised by a (positive) constant coefficient of normal restitution $\al \leq 1$. For elastic collisions $\al=1$ while $\al<1$ for inelastic collisions.
The granular gas is immersed in a \textit{molecular} gas consisting of hard disks or spheres of mass $m_g$ and diameter $\sigma_g$. The collisions between the granular particles and the gas molecules are assumed to be elastic. As mentioned in Sec.\ \ref{sec1}, we also assume that the number density of the granular particles is much smaller than that of the molecular gas, so that the state of the latter is not significantly affected by the presence of grains. In this sense, the molecular gas can be treated as a thermostat or bath in equilibrium at the temperature $T_g$. Thus, its velocity distribution function $f_g(\mathbf{V}_g)$ is
\beq
\label{1.1}
f_g(\mathbf{V}_g)=n_g \Big(\frac{m_g}{2\pi T_g}\Big)^{d/2} \exp \Bigg(-\frac{m_g V_g^2}{2T_g}\Bigg),
\eeq
where $n_g$ is the number density of molecular gas and $\mathbf{V}_g=\mathbf{v}-\mathbf{U}_g$. In principle, the mean flow velocity of molecular gas $\mathbf{U}_g$ is different from the mean flow velocity $\mathbf{U}$ of solid particles (see its definition in \eqref{1.7}). In addition, for the sake of simplicity, we take the Boltzmann constant $k_\text{B}=1$ throughout the paper.

In the low-density regime, the velocity distribution function $f(\mathbf{r}, \mathbf{v}, t)$ of granular particles verifies the Boltzmann kinetic equation. Moreover, although the granular gas is sufficiently rarefied and hence the properties of the molecular (interstitial) gas can be supposed to be constant, one has to take into account the collisions among grains themselves in the kinetic equation of $f(\mathbf{r}, \mathbf{v}, t)$. Thus, in the presence of the gravitational field $\mathbf{g}$, the distribution $f$ verifies the Boltzmann equation
\beq
\label{1.2}
\frac{\partial f}{\partial t}+\mathbf{v}\cdot \nabla f+\mathbf{g}\cdot \frac{\partial f}{\partial \mathbf{v}}=J[f,f]+J_g[f,f_g].
\eeq
Here, the Boltzmann collision operator $J[f,f]$ gives the rate of change of $f$ due to \emph{inelastic} collisions among granular particles. Its explicit form is \cite[]{BP04,G19}
\beq
\label{1.3}
J[\mathbf{v}_1|f,f]=\sigma^{d-1}\int d\mathbf{v}_{2}\int d\widehat{\boldsymbol {\sigma
}}\Theta (\widehat{{\boldsymbol {\sigma }}} \cdot {\mathbf g}_{12}) (\widehat{\boldsymbol {\sigma }}\cdot {\mathbf g}_{12})
\left[\al^{-2}f({\mathbf v}_{1}'')f({\mathbf v}_{2}'')-f({\mathbf v}_{1})f({\mathbf v}_{2})\right],
\eeq
where $\mathbf{g}_{12}=\mathbf{v}_1-\mathbf{v}_2$ is the relative velocity, $\widehat{\boldsymbol {\sigma }}$ is a unit vector that join the centers of the colliding particles, and $\Theta$ is the Heaviside step function. In Eq.\ \eqref{1.3}, the double primes denote pre-collisional velocities. The relation between them and their corresponding post-collisional velocities $(\mathbf{v}_1,\mathbf{v}_2)$ is
\beq
\label{1.4}
\mathbf{v}_1''=\mathbf{v}_1-\frac{1+\al}{2\al}(\widehat{{\boldsymbol {\sigma }}} \cdot {\mathbf g}_{12})\widehat{\boldsymbol {\sigma }}, \quad
\mathbf{v}_2''=\mathbf{v}_2+\frac{1+\al}{2\al}(\widehat{{\boldsymbol {\sigma }}} \cdot {\mathbf g}_{12})\widehat{\boldsymbol {\sigma }}.
\eeq

In Eq.\ \eqref{1.2}, the Boltzmann-Lorentz operator $J_g[f,f_g]$ accounts for the rate of change of $f$ due to \emph{elastic} collisions between particles of the granular and molecular gas. Its form is \cite[]{RL77}
\beq
\label{1.5}
J_g[\mathbf{v}_1|f,f_g]=\overline{\sigma}^{d-1}\int d\mathbf{v}_{2}\int d\widehat{\boldsymbol {\sigma
}}\Theta (\widehat{{\boldsymbol {\sigma }}} \cdot {\mathbf g}_{12}) (\widehat{\boldsymbol {\sigma }}\cdot {\mathbf g}_{12})
\left[f({\mathbf v}_{1}'')f_g({\mathbf v}_{2}'')-f({\mathbf v}_{1})f_g({\mathbf v}_{2})\right],
\eeq
where $\overline{\sigma}=(\sigma+\sigma_g)/2$ and in Eq.\ \eqref{1.5} the relationship between $(\mathbf{v}_1'',\mathbf{v}_2'')$ and $(\mathbf{v}_1,\mathbf{v}_2)$ is
\beq
\label{1.6}
\mathbf{v}_1''=\mathbf{v}_1-2 \mu_g(\widehat{{\boldsymbol {\sigma }}} \cdot {\mathbf g}_{12})\widehat{\boldsymbol {\sigma }}, \quad
\mathbf{v}_2''=\mathbf{v}_2+2 \mu (\widehat{{\boldsymbol {\sigma }}} \cdot {\mathbf g}_{12})\widehat{\boldsymbol {\sigma }}.
\eeq
Here,
\beq
\label{1.6.1}
\mu_g=\frac{m_g}{m+m_g}, \quad \mu=\frac{m}{m+m_g}.
\eeq

As is customary, the effect of gravity on the properties of molecular gases under ordinary conditions is neglected in the present analysis. This approximation is justified by the fact that the influence of gravity on a molecule between successive collisions is negligible, i.e.,  $\ell \ll h$, where $\ell$ denotes the mean free path for hard spheres, and $h = v_\text{th}^2/g$ represents the characteristic length scale over which gravitational effects become significant ($v_\text{th}$ being the thermal velocity). For instance, under terrestrial conditions at room temperature, this ratio is on the order of $\ell/h \sim 10^{-11}$ \cite[]{TGS99}, thereby validating the omission of gravitational effects in the description of molecular gas behaviour.

The number density $n$, mean flow velocity $\mathbf{U}$, and granular temperature $T$ of the granular gas are defined as the first few velocity moments of $f$:
\beq
\label{1.7}
\left\{n, n \mathbf{U},d n T\right\}=\int d\mathbf{v} \left\{1, \mathbf{v}, m V^2\right\} f(\mathbf{v}),
\eeq
where $\mathbf{V}=\mathbf{v}-\mathbf{U}$ is the peculiar velocity. As said before, the difference $\Delta \mathbf{U}=\mathbf{U}-\mathbf{U}_g$ is in general different from zero \cite[]{GG22}. In fact, as we show later, $\Delta \mathbf{U}$ induces a non-vanishing contribution to the mass flux of intruders.

The macroscopic balance equations for the granular gas are obtained by multiplying Eq.\ \eqref{1.2} by $\left\{1, \mathbf{v}, m V^2\right\}$ and integrating over velocity. The result is \cite[]{GG22}
\beq
\label{1.8}
D_t n+n\nabla\cdot \mathbf{U}=0,
\eeq
\beq
\label{1.9}
\rho D_t\mathbf{U}=-\nabla \cdot \mathsf{P}+\rho \mathbf{g}+\mathcal{F}[f],
\eeq
\beq
\label{1.10}
D_tT+\frac{2}{dn}\Big(\nabla \cdot \mathbf{q}+\mathsf{P}:\nabla \mathbf{U}\Big)=- T \zeta-T \zeta_g.
\eeq
In Eqs.\ \eqref{1.8}--\eqref{1.10}, $D_t=\partial_t+\mathbf{U}\cdot \nabla$ is the material derivative and the pressure tensor $\mathsf{P}$ and the heat flux vector $\mathbf{q}$ are given, respectively, as
\beq
\label{1.11}
\mathsf{P}=\int d\mathbf{v}\; m \mathbf{V}\mathbf{V} f(\mathbf{v}), \quad \mathbf{q}=\int d\mathbf{v}\; \frac{m}{2} V^2 \mathbf{V} f(\mathbf{v}).
\eeq
The production of momentum term $\mathcal{F}[f]$ appearing in Eq.\ \eqref{1.9} is defined as
\beq
\label{1.11.1}
\mathcal{F}[f]=\int d\mathbf{v}\; m \mathbf{V}  J_g[f,f_g].
\eeq
This term is in general different from zero since the Boltzmann--Lorentz collision operator $J_g[f,f_g]$ does not conserve momentum. The form of $\mathcal{F}[f]$ can be made more explicit when one takes into account the property \cite[]{BP04,G19}
\beqa
\label{1.11.2}
\int d\mathbf{v}_1 \Psi(\mathbf{v}_1) J_g[\mathbf{v}_2|f,f_g]&=&\overline{\sigma}^{d-1}\int d\mathbf{v}_{1}\int d\mathbf{v}_{2}\int d\widehat{\boldsymbol {\sigma
}}\Theta (\widehat{{\boldsymbol {\sigma }}} \cdot {\mathbf g}_{12}) (\widehat{\boldsymbol {\sigma }}\cdot {\mathbf g}_{12}) f(\mathbf{v}_1)f_g(\mathbf{v}_2)\nonumber\\
& & \times \left[\Psi(\mathbf{v}_1')-\Psi(\mathbf{v}_1)\right],
\eeqa
where $\mathbf{v}_1'=\mathbf{v}_1-2 \mu_g(\widehat{{\boldsymbol {\sigma }}} \cdot {\mathbf g}_{12})\widehat{\boldsymbol {\sigma }}$. Using \eqref{1.11.2}, $\mathcal{F}[f]$ is
\beq
\label{1.12}
\mathcal{F}[f]=-\frac{2\pi^{(d-1)/2}}{\Gamma\Big(\frac{d+3}{2}\Big)} \frac{m m_g}{m+m_g} \overline{\sigma}^{d-1}\int d\mathbf{v}_1\int d\mathbf{v}_2\; g_{12}\mathbf{g}_{12}\; f(\mathbf{v}_1) f_g(\mathbf{v}_2).
\eeq

Finally, the partial production rates $\zeta$ and $\zeta_g$ appearing in the balance equation \eqref{1.10} are given, respectively, as
\beq
\label{1.13}
\zeta=-\frac{m}{d n T}\int d\mathbf{v}\; V^2\; J[\mathbf{v}|f,f], \quad \zeta_g=-\frac{m}{d n T}\int d\mathbf{v}\; V^2\; J_g[\mathbf{v}|f,f_g].
\eeq
While the cooling rate $\zeta$ provides the rate of change of kinetic energy of grains due to their inelastic collisions, the term $\zeta_g$ gives the transfer of kinetic energy in the collisions between the particles of the molecular and granular gases. The quantity $\zeta=0$ for elastic collisions ($\al=1$) while $\zeta_g=0$ when the particles of the molecular and granular gases are mechanically equivalent.

It is interesting at this point to note the meaning of the \textit{granular} temperature $T$. To understand it, it is important to remember that our study is limited to the so-called rapid flow regime, namely a situation where grains are subjected to a strong external excitation (e.g. vibrating or shearing walls or air-fluidised beds). In this regime, the external energy supplied to the granular gas can compensate for the energy loss due to collisions and the effects of gravity. Since in this regime the motion of the grains is quite similar to the chaotic motion of atoms or molecules in an ordinary gas, as discussed in previous works \cite[]{GG22}, it is tempting to establish a relationship between the statistical motion of the grains and some kind of temperature. In this context, as usual in the conventional kinetic theory \cite[]{CC70}, the granular temperature $T$ can be interpreted as a measure of the fluctuations of the velocities of grains with respect to its mean value $\mathbf{U}$. Since granular gases are \textit{athermal} systems (i.e., their thermal fluctuations have a negligible effect on the dynamics of grains), the granular temperature $T$ has no thermodynamic interpretation in contrast to the temperature $T_g$ of the molecular gas (see for example the review paper of \cite{G08bis} for a discussion of this issue). In any case, within the context of the statistical thermodynamics, the thermodynamic temperature $T_g$ can be also understood as a statistical quantity measuring the deviations of molecular particles's velocity $\mathbf{v}$ from its mean value $\mathbf{U}_g$.

Before closing this section, it is worth analysing the
limiting case $m\gg m_g$. It is in fact a quite realistic case for granular suspensions \cite[]{S20} where the particles of the interstitial molecular gas are much lighter than the particles of the granular gas. In the limit $m/m_g\to \infty$, a Kramers--Moyal expansion \cite[]{RL77} allows us to approximate
the Boltzmann--Lorentz operator $J_g[f,f_g]$ to the Fokker--Planck operator $J_g^{\text{FP}}[f,f_g]$:
\beq
\label{1.14}
J_g[f,f_g]\to J_g^{\text{FP}}[f,f_g]=\gamma \frac{\partial}{\partial \mathbf{v}}\cdot \Bigg(\mathbf{v}+\frac{T_g}{m}\frac{\partial}{\partial \mathbf{v}}\Bigg)f(\mathbf{v}),
\eeq
where the friction coefficient $\gamma$ is
\beq
\label{1.15}
\gamma=\frac{4\pi^{(d-1)/2}}{d\Gamma\Big(\frac{d}{2}\Big)}\Big(\frac{m_g}{m}\Big)^{1/2}\Bigg(\frac{2T_g}{m}\Bigg)^{1/2}n_g \overline{\sigma}^{d-1}.
\eeq
Upon deriving Eq.\ \eqref{1.14}, it has been assumed that $\mathbf{U}_g=\mathbf{0}$ and that the distribution function $f$ of the granular gas is a Maxwellian distribution.

Most of the theoretical works for suspension models reported in the granular literature are essentially based on the use of the Fokker--Planck operator \eqref{1.14} to account for in an effective way (coarse-grained approach) the influence of the surrounding fluid on the dynamics of grains \cite[]{KH01,GTSH12,ChBCh23,G23}. This sort of models have been considered to obtain the Navier--Stokes--Fourier transport coefficients of the suspension \cite[]{GGG19a}.

\section{Intruders in granular suspensions}
\label{sec3}

We assume now that a few intruders (or tracers) of mass $m_0$ and diameter $\sigma_0$ are added to the granular gas. In this situation, intruders and particles of the granular gas are surrounded by the molecular gas (bath of elastic hard spheres).
The system can be seen as a ternary mixture where one of the components (intruders) are present in tracer concentration. Apart from the restitution coefficient $\al$ for inelastic grain-grain collisions, the coefficient of normal restitution $\al_0\leq 1$ characterizes the inelastic collisions between the intruders and the particles of the granular gas. As in the case of the granular gas, collisions between intruders and particles of the surrounding molecular gas are elastic.

Since the concentration of intruders is much smaller than that of the granular gas (tracer limit), its presence does not affect the state of the granular gas. Under these conditions and in the presence of the gravitational field, the velocity distribution function $f_0(\mathbf{r}, \mathbf{v}; t)$ of the intruders obeys the kinetic equation
\beq
\label{2.1}
\frac{\partial f_0}{\partial t}+\mathbf{v}\cdot \nabla f_0+\mathbf{g}\cdot \frac{\partial f_0}{\partial \mathbf{v}}=J_0[f_0,f]+J_{0g}[f_0,f_g],
\eeq
where the (inelastic) version of the Boltzmann--Lorentz collision operator $J_0[f_0,f]$ gives the rate of change of $f_0$ due to the \emph{inelastic} collisions between the intruders and particles of the granular gas. It is given  by \cite[]{G19}
\beq
\label{2.2}
J_0[\mathbf{v}_1|f_0,f]=\sigma^{'(d-1)}\int d\mathbf{v}_{2}\int d\widehat{\boldsymbol {\sigma
}}\Theta (\widehat{{\boldsymbol {\sigma}}} \cdot {\mathbf g}_{12}) (\widehat{\boldsymbol {\sigma }}\cdot {\mathbf g}_{12})
\left[\al_0^{-2}f_0({\mathbf v}_{1}'')f({\mathbf v}_{2}'')-f_0({\mathbf v}_{1})f({\mathbf v}_{2})\right].
\eeq
As in Eq.\ \eqref{1.3}, $\mathbf{g}_{12}=\mathbf{v}_1-\mathbf{v}_2$ is the relative velocity, $\widehat{\boldsymbol {\sigma }}$ is a unit vector, and $\Theta$ is the Heaviside step function. In addition, $\sigma'=(\sigma+\sigma_0)/2$,
\beq
\label{2.3}
\mathbf{v}_1''=\mathbf{v}_1-\frac{1+\al_0}{\al_0}\mu'(\widehat{{\boldsymbol {\sigma }}} \cdot {\mathbf g}_{12})\widehat{\boldsymbol {\sigma }}, \quad
\mathbf{v}_2''=\mathbf{v}_2+\frac{1+\al_0}{2\al_0}\mu_0'(\widehat{{\boldsymbol {\sigma }}} \cdot {\mathbf g}_{12})\widehat{\boldsymbol {\sigma}},
\eeq
and
\beq
\label{2.3.1}
\mu'=\frac{m}{m+m_0}, \quad  \mu_0'=\frac{m_0}{m+m_0}.
\eeq

In Eq.\ \eqref{2.1}, the collision operator $J_{0g}[f_0,f_g]$ provides the rate of change of $f_0$ due to \emph{elastic} collisions between intruders and particles of the molecular gas. Similarly to the operator $J_{0}[f,f_g]$, it is given by
\beq
\label{2.4}
J_{0g}[\mathbf{v}_1|f_0,f_g]=\overline{\sigma}_0^{d-1}\int d\mathbf{v}_{2}\int d\widehat{\boldsymbol {\sigma
}}\Theta (\widehat{{\boldsymbol {\sigma }}} \cdot {\mathbf g}_{12}) (\widehat{\boldsymbol {\sigma }}\cdot {\mathbf g}_{12})
\left[f_0({\mathbf v}_{1}'')f_g({\mathbf v}_{2}'')-f_0({\mathbf v}_{1})f_g({\mathbf v}_{2})\right].
\eeq
Here, $\overline{\sigma}_0=(\sigma_0+\sigma_g)/2$,
\beq
\label{2.5}
\mathbf{v}_1''=\mathbf{v}_1-2 \mu_{g0}(\widehat{{\boldsymbol {\sigma }}} \cdot {\mathbf g}_{12})\widehat{\boldsymbol {\sigma }}, \quad
\mathbf{v}_2''=\mathbf{v}_2+2 \mu_{0g} (\widehat{{\boldsymbol {\sigma }}} \cdot {\mathbf g}_{12})\widehat{\boldsymbol {\sigma }}.
\eeq
and
\beq
\label{2.5.1}
\mu_{g0}=\frac{m_g}{m_0+m_g}, \quad  \mu_{0g}=\frac{m_0}{m_0+m_g}.
\eeq

Although the granular temperature $T$ is the relevant one at a hydrodynamic level, an interesting quantity at a kinetic level is the local temperature of the intruders $T_0$. This quantity measures the mean kinetic energy of the intruders. It is defined as
\begin{equation}
\label{2.5.2}
T_0({\bf r}, t)=\frac{m_0}{d n_0({\bf r}, t)}\int \; d{\bf v}\, V^2 f_0({\bf r},{\bf v},t).
\end{equation}
As confirmed by kinetic theory calculations \cite[]{GD99b}, computer simulations \cite[]{G19}, and experiments \cite[]{WP02,FM02,PTHS24} the global temperature $T$ and the temperature of impurities $T_0$ are in general different. 

Intruders may freely exchange momentum and energy with the particles of the granular and molecular gas. Thus, only the number density of intruders
\beq
\label{2.6}
n_0(\mathbf{r}; t)=\int d\mathbf{v} f_0(\mathbf{r}, \mathbf{v}, t)
\eeq
is conserved. This yields the balance equation
\beq
\label{2.7}
\frac{\partial \rho_0}{\partial t}+\nabla \cdot \mathbf{j}_0=0,
\eeq
where $\rho_0=m_0 n_0$ is the mass density of intruders and
\beq
\label{2.8}
\mathbf{j}_0(\mathbf{r}; t)= \int d\mathbf{v}\; m_0 \mathbf{V}\; f_0(\mathbf{r}, \mathbf{v};t)
\eeq
is the mass flux of intruders.

As in the case of the Boltzmann--Lorentz operator $J_g[f,f_g]$, in the limiting case $m_0\gg m_g$ the operator $J_{0g}[f_0,f_g]$ reduces to the Fokker--Planck operator
\beq
\label{2.9}
J_{0g}[f_0,f_g]\to J_{0g}^{\text{FP}}[f_0,f_g]=\gamma_0 \frac{\partial}{\partial \mathbf{v}}\cdot \Bigg(\mathbf{v}+\frac{T_g}{m_0}\frac{\partial}{\partial \mathbf{v}}\Bigg)f_0(\mathbf{v}),
\eeq
where the friction coefficient $\gamma_0$ is
\beq
\label{2.10}
\gamma_0=\frac{4\pi^{(d-1)/2}}{d\Gamma\Big(\frac{d}{2}\Big)}\Big(\frac{m_g}{m_0}\Big)^{1/2}\Bigg(\frac{2T_g}{m_0}\Bigg)^{1/2}n_g \overline{\sigma}_0^{d-1}.
\eeq
Note that the expression \eqref{2.10} of $\gamma_0$  differs from the macroscopic Stokes law describing the Brownian motion of a massive particle in an equilibrium host fluid 
\cite[]{BPH94,BHP94,GG22a}. This means that the results derived in this paper in the Brownian limit does not exactly reduce to those obtained by using a coarse-grained approach \cite[]{GG22a}.

\section{Homogeneous steady state}
\label{sec4}

Before considering inhomogeneous situations, it is pertinent to analyse the homogeneous steady state (HSS) for the system (intruders and granular gas immersed in a molecular gas) in the absence of the gravitational field ($\mathbf{g}=\mathbf{0}$). The study of this state for the intruders is crucial since the HSS plays the role of the reference state in the Chapman--Enskog method \cite[]{CC70}.

In the HSS state, the densities $n$ and $n_0$ and the granular temperature $T$ are spatially uniform. Moreover, without loss of generality, the mean flow velocities vanish ($\mathbf{U}=\mathbf{U}_g=\mathbf{0}$) with an appropriate choice of the frame reference. Under these conditions, Eq.\ \eqref{1.2} for the granular gas reads
\beq
\label{3.1}
\partial_t f=J[f,f_0]+J_{g}[f,f_g],
\eeq
while Eq.\ \eqref{2.1} for the intruders becomes
\beq
\label{3.2}
\partial_t f_0=J_0[f_0,f]+J_{0g}[f_0,f_g].
\eeq

\subsection{HSS for the granular gas}

Given that the HSS for the granular gas was already analysed by \cite{GG22}, only some few results are provided here.
Since in the HSS the distribution $f(\mathbf{v})$ is isotropic in $\mathbf{v}$, then $\mathcal{F}[f]=\mathbf{0}$. On the other hand, the kinetic energy is not conserved by collisions and so $\zeta\neq 0$ and $\zeta_g \neq 0$. The most interesting quantity in the HSS for the granular gas is the temperature ratio $\chi=T/T_g$, which is in general different from 1. The only nontrivial balance equation in the homogenous state for the granular gas is that of the temperature \eqref{1.10}:
\beq
\label{3.3}
\frac{\partial T}{\partial t}=-T\left(\zeta+\zeta_g\right).
\eeq
After a transient period, it is expected that the granular gas achieves a steady state. Thus, according to Eq.\ \eqref{3.3}, the steady-state condition is $\zeta+\zeta_g=0$. As discussed by \cite{GG22}, as the molecular gas acts as a thermostat in the steady state, then the mean kinetic energy of the granular particles is smaller than that of the molecular gas ($T<T_g$). This necessarily requires that $\zeta_g<0$ so that, in the steady state the production rates $\zeta$ and $|\zeta_g|$ exactly compensate each other and one achieves the condition $\zeta+\zeta_g=0$.


However, to determine $\zeta$ and $\zeta_g$ one needs to know the velocity distribution function $f(\mathbf{v})$ for the granular gas. For inelastic collisions ($\al<1$) this distribution is not exactly known to date. On the other hand, the results obtained for the  fourth cumulant or \textit{kurtosis} $c$ of the distribution $f$ in \cite{GG22} clearly show (see Fig.\ 3 of \cite{GG22}) that the magnitude of $c$ is in general very small for not quite strong inelasticity (e.g. $\al \gtrsim 0.5$). Thus, to estimate the production rates $\zeta$ and $\zeta_g$ one can replace the true distribution $f(\mathbf{v})$ by the Maxwellian distribution 
\beq
\label{3.5}
f_\text{M}(\mathbf{v})=n \Big(\frac{m}{2\pi T}\Big)^{d/2} \exp \Big(-\frac{m v^2}{2T}\Big).
\eeq
In the Maxwellian approximation, the dimensionless production rates $\zeta^*=\zeta/\nu$ and $\zeta_g^*=\zeta_g/\nu$ are given by \cite[]{GG22}
\beq
\label{3.6}
\zeta^*=\frac{\sqrt{2}\pi^{(d-1)/2}}{d\Gamma\Big(\frac{d}{2}\Big)}(1-\al^2),
\quad \zeta_g^*=2x(1-x^2) \Big(\frac{\mu T}{T_g}\Big)^{1/2}\gamma^*,
\eeq
where $\nu=n\sigma^{d-1}\sqrt{2T/m}$ is an effective collision frequency,
\beq
\label{3.8}
x=\Bigg(\mu_g+\mu \frac{T_g}{T}\Bigg)^{1/2},
\eeq
and
\beq
\label{3.9}
\gamma^*=\epsilon \; \chi^{-1/2}, \quad \epsilon=\frac{\ell \gamma}{\sqrt{2T_g/m}}=\frac{\sqrt{2}\pi^{d/2}}{2^d d \Gamma\left(\frac{d}{2}\right)}\frac{1}{\phi \sqrt{T_g^*}}.
\eeq
Here, $\ell=1/n\sigma^{d-1}$ is proportional to the mean free path of hard spheres,
\beq
\label{3.10}
\phi=\frac{\pi^{d/2}}{2^{d-1}d \Gamma\left(\frac{d}{2}\right)} n\sigma^d
\eeq
is the solid volume fraction and
\beq
\label{3.11}
T_g^*=\frac{T_g}{m\sigma^2 \gamma^2}.
\eeq

In the Maxwellian approximation (i.e, when one replaces $f$ by $f_\text{M}$), the steady temperature ratio $T/T_g$ can be obtained by inserting the expressions \eqref{3.6}  of $\zeta^*$  and $\zeta_g^*$, respectively, into the (exact) steady-state condition $\zeta^*+\zeta_g^*=0$.  This yields a cubic equation for the quantity $x$ whose physical solution is given by Eq.\ (4.14) of the supplementary material of \cite{GG22}. In terms of $x$, the final expression of the temperature ratio $T/T_g$ is given by Eq.\ (4.15) of the above supplementary material. In spite of considering the Maxwellian approximation for the distribution $f$, an excellent agreement between theory and simulations for the temperature ratio is observed over the whole range of values of $\al$ studied (see figure 1 of \cite{GG22}).   

\subsection{HSS for the intruders}

We analyse now the HSS for the intruders. The
balance equation for the intruders' temperature $T_0$ can be easily obtained from Eq.\ \eqref{3.2} as $\partial_t \ln T_0=-\left(\zeta_0+\zeta_{0g}\right)$,
where
\beq
\label{3.16}
\zeta_0=-\frac{m_0}{d n_0 T_0}\int d\mathbf{v}\; v^2\; J_0[\mathbf{v}|f_0,f], \quad
\zeta_{0g}=-\frac{m_0}{d n_0 T_0}\int d\mathbf{v}\; v^2\; J_{0g}[\mathbf{v}|f_0,f_g].
\eeq
In the HSS, $\partial_t T_0=0$ and the condition for obtaining $T_0$ is
\beq
\label{3.16.1}
\zeta_0+\zeta_{0g}=0.
\eeq

As in the case of the granular gas, the exact form of the distribution function $f_0(\mathbf{v})$ for inelastic collisions is not known to date. The departure of $f_0(\mathbf{v})$ from its Maxwellian form
\beq
\label{3.17}
f_{0,\text{M}}(\mathbf{v})=n_0 \Big(\frac{m_0}{2\pi T_0}\Big)^{d/2} \exp \Big(-\frac{m_0 v^2}{2T_0}\Big)
\eeq
can be measured by the kurtosis $c_0$. It is defined as \cite[]{G19}
\beq
\label{3.17.1}
c_0=\frac{1}{d(d+2)}\frac{m_0^2}{n_0T_0^2}\int d\mathbf{v}\; v^4 f_0(\mathbf{v})-1.
\eeq
\begin{figure}
\begin{center}
\includegraphics[width=0.55\textwidth]{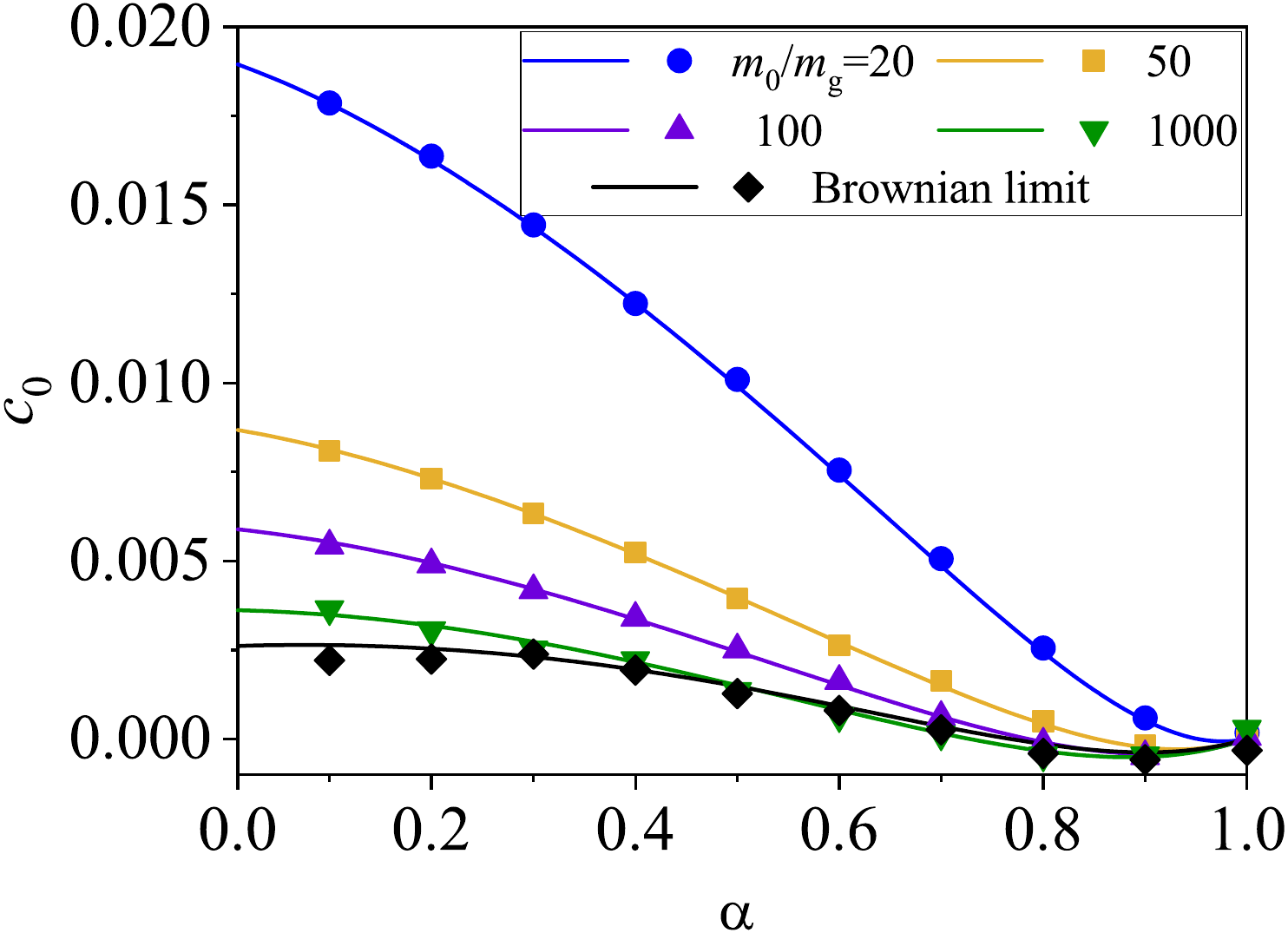}
\caption{Plot of the kurtosis $c_0$ associated with the distribution function of the intruders as a function of the coefficient of normal restitution $\al$ for $d=3$, $\phi=0.0052$, $T_g^*=1000$, and four different values of the mass ratio $m_0/m_g$ [from top to bottom, $m_0/m_g=20, 50, 100$, and 1000]. Moreover, in all the curves $m_0/m=10$, $\sigma_0/\sigma=5$, and $\sigma_0/\sigma_g=(m_0/m_g)^{1/3}$. The solid lines are the theoretical results while the symbols are the DSMC simulation results. The dashed line is the result obtained from the Fokker--Planck approach \eqref{2.9} to the operator $J_{0g}[f_0,f_g]$. \vicente{Diamonds} refer to DSMC simulations implemented using the time-driven approach \cite[]{GG22}.}
\label{fig3}
\end{center}
\end{figure}

Some technical details on the determination of $c_0$ are given in the supplementary material. Given that the expression of $c_0$ is very large and not very illuminating, its final form is not displayed here. In terms of dimensionless quantities, the parameter space of a $d$-dimensional system is given by the set
\beq
\label{3.18}
\xi\equiv \left\{\frac{\sigma}{\sigma_g}, \frac{\sigma_0}{\sigma_g}, \frac{m}{m_g}, \frac{m_0}{m_g}, \al, \al_0,  \phi, T_g^*\right\}.
\eeq
In contrast to the monocomponent case \cite[]{GG22}, note that the diameter ratios $\sigma/\sigma_g$ and $\sigma_0/\sigma_g$ appear also as input parameters of the system.    

Figure \ref{fig3} shows $c_0$ versus the (common) coefficient of restitution $\al=\al_0$ for $d=3$, $\phi=0.0052$, and $T_g^*=1000$. Four different values of the mass ratio $m_0/m_g$ ($m_0/m_g=20, 50, 100$, and 1000) are considered keeping the ratio $m_0/m=10$. In addition, $\sigma_0/\sigma=5$ and we have assumed that the intruders and molecular gas particles have the same mass density [i.e., $\sigma_0/\sigma=(m_0/m_g)^{1/3}$]. As occurs for the kurtosis $c$ of the granular gas \vicente[{see figure 3 of \cite{GG22}]}, it is quite apparent from figure \ref{fig3} that the magnitude of $c_0$ is in general quite small. \vicente{We observe that} the agreement between theory and DSMC simulations is excellent even for quite extreme values of inelasticity.

The fact that $c_0$ is small allows us to guarantee that a good estimate of the production rates $\zeta_0$ and $\zeta_{0g}$ can be obtained by replacing $f_0(\mathbf{v})$ by the Maxwellian distribution $f_{0,\text{M}}(\mathbf{v})$ in Eq.\ \eqref{3.16}. In this approximation,
the dimensionless quantities $\zeta_0^*=\zeta_0/\nu$ and $\zeta_{0g}^*=\zeta_{0g}/\nu$ can be written as
\beq
\label{3.19}
\zeta_0^*=\frac{4\pi^{(d-1)/2}}{d\Gamma\left(\frac{d}{2}\right)}\mu' \left(\frac{\sigma'}{\sigma}\right)^{d-1}\left(1+\frac{m T_0}{m_0 T}\right)^{1/2}(1+\al_0)\left[1-\frac{\mu'}{2}(1+\al_0)\left(1+\frac{m_0 T}{m T_0}\right)\right],
\eeq
\beq
\label{3.20}
\zeta_{0g}^*=2x_0(1-x_0^2)\left(\mu_{0g}\frac{T_0}{T_g}\right)^{1/2}\gamma_0^*.
\eeq
In Eqs.\ \eqref{3.19} and \eqref{3.20}, we have introduced the quantities
\beq
\label{3.21}
x_0=\left(\mu_{g0}+\mu_{0g}\frac{T_g}{T_0}\right)^{1/2},
\eeq
\beq
\label{3.21.1}
\gamma_0^*=\epsilon_0\chi^{-1/2}, \quad
\epsilon_0=
\left(\frac{\overline{\sigma}_0}{\overline{\sigma}}\right)^{d-1}\frac{m}{m_0}\epsilon.
\eeq
The temperature ratio $\chi_0\equiv T_0/T_g$ can be finally determined by substituting Eqs.\ \eqref{3.19} and \eqref{3.20} into the condition \eqref{3.16.1}. It yields the nonlinear algebraic equation
\beqa
\label{3.22}
2x_0(x_0^2-1)\left(\mu_{0g}\chi_0\right)^{1/2}\gamma_0^*&=&\frac{4\pi^{(d-1)/2}}{d\Gamma\left(\frac{d}{2}\right)}\mu' \left(\frac{\sigma'}{\sigma}\right)^{d-1}\left(1+\frac{m \chi_0}{m_0 \chi}\right)^{1/2}(1+\al_0)\nonumber\\
& & \times \left[1-\frac{\mu}{2}(1+\al_0)\left(1+\frac{m_0 \chi}{m \chi_0}\right)\right],
\eeqa
where we recall that $\chi=T/T_g$ is given by Eq.\ (3.7) of the supplementary material of \cite{GG22}.
The numerical solution to Eq.\ \eqref{3.21} provides the dependence of $T_0/T_g$ on the parameter space $\xi$.

\begin{figure}
\begin{center}
\includegraphics[width=0.55\textwidth]{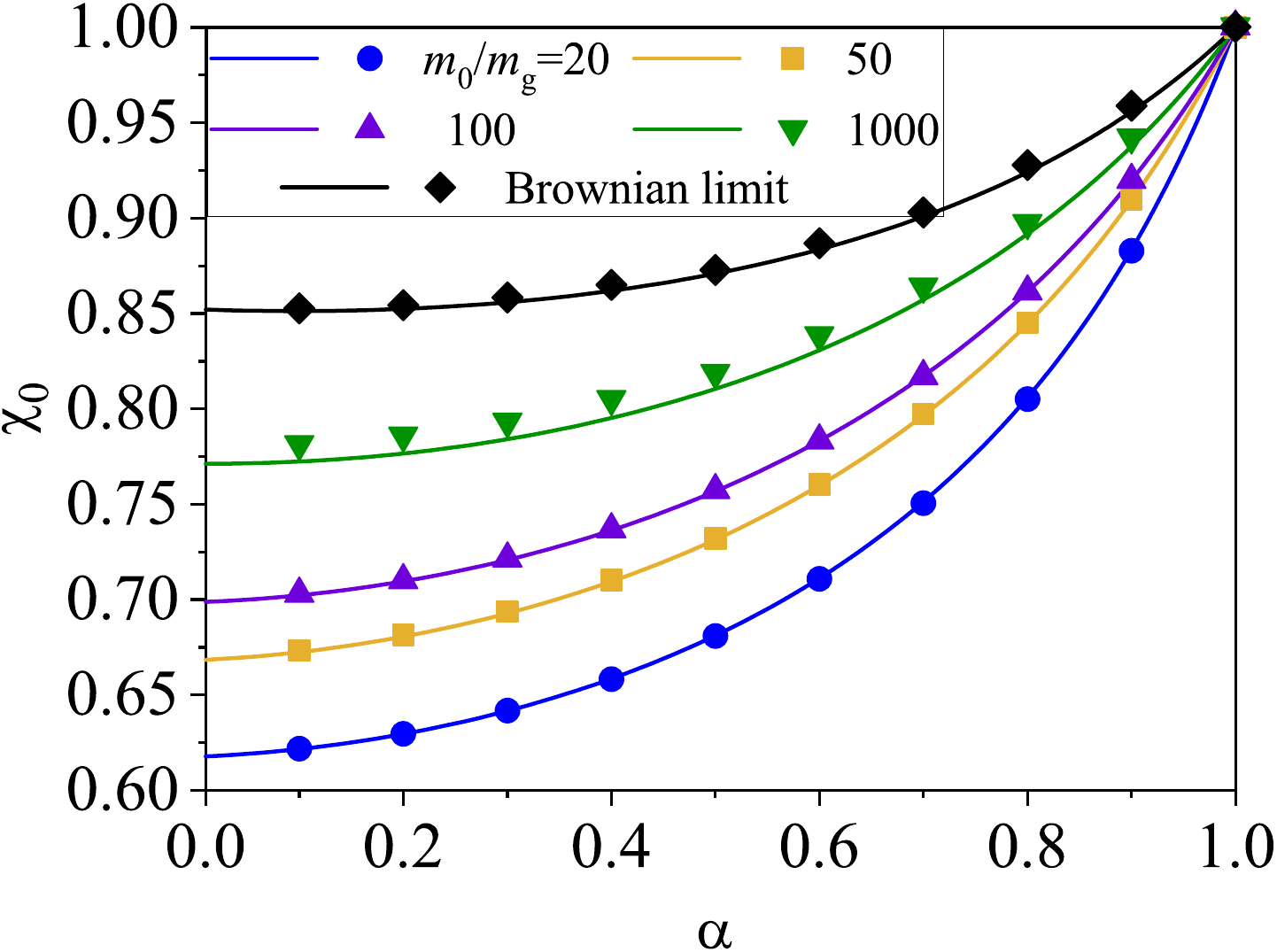}
\caption{Temperature ratio $\chi_0\equiv T_0/T_g$ versus the (common) coefficient of normal restitution $\al_0=\al$ for $d=3$, $\phi=0.0052$, $T_g^*=1000$, and four different values of the mass ratio $m_0/m_g$ [from top to bottom, $m_0/m_g=20, 50, 100$, and 1000]. Moreover, in all the curves $m_0/m=10$, $\sigma_0/\sigma=5$, and $\sigma_0/\sigma_g=(m_0/m_g)^{1/3}$. The solid lines are the theoretical results while the symbols are the Monte Carlo simulation results. The dashed line is the result obtained by using the Fokker--Planck approach \eqref{2.9} to the operator $J_{0g}[f_0,f_g]$. \vicente{Diamonds} refer to DSMC simulations implemented using the time-driven approach \cite[]{GG22}.}
\label{fig4}
\end{center}
\end{figure}

For the sake of illustration, figure \ref{fig4} shows $\chi_0\equiv T_0/T_g$ as a function of the (common) coefficient of restitution $\al_0=\al$ for the same systems as in figure \ref{fig3}. As occurs for the ratio $T/T_g$, due to the way of scaling the relevant quantities of the system, the deviation of $\chi_0$ from unity increases with decreasing the mass ratio $m_0/m_g$. The agreement between the (approximate) theoretical results and computer simulations is again excellent; it clearly justifies the use of the Maxwellian distribution \eqref{3.17} to achieve accurate estimates of the cooling rates $\zeta_0$ and $\zeta_{0g}$.

A point to consider here is that convergence to the results obtained using the Fokker-Planck model is achieved not only in the limit $m/m_g\to\infty$ and $m_0/m_g\to\infty$ but it is also necessary that $\sigma/\sigma_g\to\infty$ and $\sigma_0/\sigma_g\to\infty$. For convenience, we will assume in the rest of the work that $m/m_g\to\infty$ and $m_0/m_g\to\infty$ also imply $\sigma/\sigma_g\to\infty$ and $\sigma_0/\sigma_g\to\infty$, and thus intruders and molecular gas particles have the same particle mass density (i.e, $m_0/\sigma_0^d=m_g/\sigma_g^d$).

\section{Chapman--Enskog method. Diffusion transport coefficients}
\label{sec5}

We assume that we perturb the homogeneous state by small spatial gradients. These perturbations induce non-vanishing contributions to the mass, momentum, and heat fluxes. The determination of these fluxes allow us to identify the corresponding Navier--Stokes--Fourier transport coefficients of the granular suspension. As said in Sec.\ \ref{sec1}, since in the tracer limit the pressure tensor and the heat flux vector of the binary mixture (intruders plus granular gas) are the same as that for the excess species (granular gas), the mass transport of intruders $\mathbf{j}_0$ is the relevant flux of the problem. The Navier--Stokes--Fourier transport coefficients of the granular gas were already determined by \cite{GG22}.

To get the mass flux $\mathbf{j}_0$, the Boltzmann kinetic equation \eqref{2.1} is solved up to first order in spatial gradients by means of the
Chapman--Enskog expansion \cite[]{CC70} conveniently adapted to dissipative dynamics. As widely discussed in many textbooks, \cite[]{CC70,FK72} there are two different stages in the relaxation of a molecular gas towards equilibrium. For times of the order of the mean free time, one can identify a first stage (kinetic stage) where the main effect of collisions on the distribution function is to relax it towards the so-called \textit{local} equilibrium state. Then, a \textit{hydrodynamic} (slow) stage is achieved where the gas has completely forgotten its initial preparation. In this stage, the microscopic state of the gas is completely specified by the knowledge of the hydrodynamic fields (in the case of a binary mixture by $n_0$, $n$, $\mathbf{U}$, and $T$). The above two stages are also expected in the case of granular gases except that in the kinetic stage the distribution function will generally relax towards a time-dependent nonequilibrium distribution (the homogeneous cooling state for freely cooling dry granular gases) instead of the local equilibrium distribution. A crucial point is that although the granular temperature $T$ is not a conserved field (due to the inelastic character of the collisions), it is assumed that $T$ can still be considered as a slow field. This assumption has been clearly supported by the good agreement found between granular hydrodynamics and computer simulations in several non-equilibrium situations \cite[]{LBD02,DHGD02,LLC07,MDCPH11,ChS13,MGH14}. More details on the application of the Chapman--Enskog method to \textit{dry} (no gas phase) granular mixtures can be found, for example, in \cite{G19}.

Based on the above arguments, in the hydrodynamic regime, the kinetic equation \eqref{2.1} admits a \emph{normal} (or hydrodynamic) solution where all the space and time dependence of $f_0$ only occurs through a functional dependence on the hydrodynamic fields. As usual \cite[]{CC70}, this functional dependence can be made explicit by assuming small spatial gradients. In this case, $f_0$ can be written as a series expansion in powers of the spatial gradients of the hydrodynamic fields:
\beq
\label{4.1}
f_0=f_0^{(0)}+f_0^{(1)}+\cdots,
\eeq
where the approximation $f_0^{(k)}$ is of order $k$ in the spatial gradients. In addition, in the presence of the gravitational force, it is necessary to characterize the magnitude of the force relative to that of the spatial gradients. Here, as in the case of the conventional fluid mixtures \cite[]{CC70}, we assume that the magnitude of $\mathbf{g}$ is at least of first order in the perturbation expansion. The implementation of the Chapman--Enskog method to first order in the spatial gradients follows similar steps as those made in the conventional inelastic hard sphere model for dry granular mixtures \cite[]{GD02,GDH07,GHD07}.

In contrast to the application of the Chapman--Enskog method for dry granular mixtures, although we are interested here in calculating the diffusion transport coefficients in steady states, the presence of the surrounding molecular gas yields in inhomogeneous states a local energy unbalance between the energy supplied by the bath (or thermostat) and the energy lost by inelastic collisions. This means that we must first consider a time-dependent reference distribution $f^{(0)}(\mathbf{r}, \mathbf{v}, t)$ to obtain the time-dependent linear integral equations that verify the diffusion coefficients. One then assumes stationary conditions and solves (approximately) the above integral equations by considering the so-called first-Sonine approximation.
In addition, as discussed by \cite{GG22}, the term $\Delta \mathbf{U}=\mathbf{U}-\mathbf{U}_g$ must be considered to be at least of first order in spatial gradients. In this case, the Maxwellian distribution $f_g(\mathbf{r},\mathbf{v}, t)$ can be written as
\beq
\label{4.2}
f_g(\mathbf{v})=f_g^{(0)}(\mathbf{V})+f_g^{(1)}(\mathbf{V})+\cdots,
\eeq
where
\beq
\label{4.3}
f_g^{(0)}(\mathbf{V})=n_g \Big(\frac{m_g}{2\pi T_g}\Big)^{d/2} \exp \Bigg(-\frac{m_g V^2}{2T_g}\Bigg), \quad 
f_g^{(1)}(\mathbf{V})=-\frac{m_g}{T_g}\mathbf{V}\cdot \Delta \mathbf{U}f_g^{(0)}(\mathbf{V}).
\eeq

The mathematical steps involved in the determination of the zeroth- and first-order distribution functions are quite similar to those made in previous works on granular mixtures \cite[]{GD02,GM07}. Technical details carried out in this derivation are provided in the supplementary material. In particular, the first-order distribution function $f^{(1)}\mathbf{r}, \mathbf{v}, t)$ is given by
\beqa
\label{4.5}
f_0^{(1)}(\mathbf{V})&=&\boldsymbol{\mathcal{A}}_0(\mathbf{V})\cdot \nabla T+\boldsymbol{\mathcal{B}}_0(\mathbf{V})\cdot \nabla n+\boldsymbol{\mathcal{C}}_0(\mathbf{V})\cdot \nabla n_0+\mathcal{D}_0'(\mathbf{V}) \nabla \cdot \mathbf{U}\nonumber\\
& &+\mathcal{D}_{0,ij}(\mathbf{V})\left(\partial_iU_j+\partial_j U_i-\frac{2}{d}\delta_{ij}\nabla\cdot \mathbf{U}\right)
+\boldsymbol{\varepsilon}_0(\mathbf{V})\cdot \Delta \mathbf{U}.
\eeqa
where the unknowns $(\boldsymbol{\mathcal{A}}_0, \boldsymbol{\mathcal{B}}_0, \boldsymbol{\mathcal{C}}_0, \mathcal{D}_{0,ij},\mathcal{D}_0',\boldsymbol{\varepsilon}_0)$ are the solutions of a set of coupled linear integral equations displayed in the supplementary material.

\subsection{Diffusion transport coefficients}

The constitutive equation for the mass flux $\mathbf{j}_0^{(1)}$ is given by Eq.\ \eqref{0.1}. The diffusion transport coefficients are defined as
\beq
\label{4.6}
D_T=-\frac{m_0 T}{d\rho}\int d\mathbf{v}\; \mathbf{V}\cdot \boldsymbol{\mathcal{A}}_0(\mathbf{V}),
\eeq
\beq
\label{4.7}
D=-\frac{n}{d}\int d\mathbf{v}\; \mathbf{V}\cdot \boldsymbol{\mathcal{B}}_0(\mathbf{V}),
\eeq
\beq
\label{4.8}
D_0=-\frac{\rho}{dm_0}\int d\mathbf{v}\; \mathbf{V}\cdot \boldsymbol{\mathcal{C}}_0(\mathbf{V}),
\eeq
\beq
\label{4.9}
D_0^U=-\frac{m_0}{d}\int d\mathbf{v}\; \mathbf{V}\cdot \boldsymbol{\mathcal{\varepsilon}}_0(\mathbf{V}).
\eeq

The procedure for obtaining the expressions of the set of coefficients $(D_T, D, D_0, D_0^U)$ is described in the the supplementary material and only their final forms are provided in section \ref{sec6}. It is important to recall that the expressions of the diffusion coefficients cannot be analytically obtained for general unsteady conditions since it would require to numerically solve a set of coupled differential equations for them. Thus, to reach analytical expressions for these diffusion coefficients, one assumes the validity of the steady-state constraint $\zeta+\zeta_g=0$ at each point of the system. This allows us to achieve explicit forms for the set $(D_T, D, D_0, D_0^U)$.


\section{Sonine polynomial approximation to the diffusion transport coefficients in steady-state conditions}
\label{sec6}

As mentioned before, the diffusion transport coefficients are given in terms of the solutions of a set of coupled linear integral equations. As usual in kinetic theory of both molecular and granular gases, these integral equations can be approximately solved by considering the leading terms of a Sonine polynomial expansion of the unknowns $\boldsymbol{\mathcal{A}}_0$, $\boldsymbol{\mathcal{B}}_0$, $\boldsymbol{\mathcal{C}}_0$, and $\boldsymbol{\varepsilon}_0$. In the case of the mass flux, the above quantities are approximated by the polynomials
\beq
\label{6.1}
\boldsymbol{\mathcal{A}}_0(\mathbf{V})\to -f_{0,\text{M}}\mathbf{V}\frac{\rho}{Tn_0T_0}D_T, \quad
\boldsymbol{\mathcal{B}}_0(\mathbf{V})\to -f_{0,\text{M}}\mathbf{V}\frac{m_0}{n n_0T_0}D,
\eeq
\beq
\label{6.2}
\boldsymbol{\mathcal{C}}_0(\mathbf{V})\to -f_{0,\text{M}}\mathbf{V}\frac{m_0^2}{\rho n_0T_0}D_0, \quad
\boldsymbol{\mathcal{\varepsilon}}_0(\mathbf{V})\to -f_{0,\text{M}}\mathbf{V}\frac{D_0^U}{n_0T_0}.
\eeq
To determine $D_0$, $D$, $D_T$, and $D_0^U$, we substitute first $\boldsymbol{\mathcal{A}}_0$, $\boldsymbol{\mathcal{B}}_0$, $\boldsymbol{\mathcal{C}}_0$, and $\boldsymbol{\varepsilon}_0$ by their leading Sonine approximations \eqref{6.1} and \eqref{6.2} in the corresponding integral equations. Then we multiply these equations by $m_0\mathbf{V}$ and integrate over velocity. Technical details on these calculations are displayed in the supplementary material.

\subsection{Thermal diffusion coefficient $D_T$}

The thermal diffusion coefficient $D_T$ is given by
\beq
\label{6.3}
D_T=\frac{n_0 T}{\rho \nu}D_T^*, \quad
D_T^*=\frac{\tau_0-\frac{m_0}{m}+\chi\frac{\partial \tau_0}{\partial \chi}}{\beta \gamma^*+\nu_D^*+\widetilde{\nu}_D\gamma_0^*},
\eeq
where $\gamma_0^*$ is defined by Eq.\ \eqref{3.21.1}, $\tau_0=T_0/T$, and
\beq
\label{6.4}
\beta=\left(x^{-1}-3x\right)\mu^{3/2}\chi^{-1/2}.
\eeq
Here, $x$ is given by Eq.\ \eqref{3.8} and in Eq.\ \eqref{6.3} we have introduced the (reduced) collision frequencies
\beq
\label{6.5}
\nu_D^*=\frac{2\pi^{(d-1)/2}}{d\Gamma\left(\frac{d}{2}\right)}\left(\frac{\sigma'}{\sigma}\right)^{d-1}\mu' \left(\frac{1+\theta_0}{\theta_0}\right)^{1/2}(1+\al_0),
\eeq
\beq
\label{6.6}
\widetilde{\nu}_D=\left(\frac{m_0T_0}{m_g T_g}\right)^{1/2}\mu_{g0}\left(1+\theta_{0g}\right)^{1/2},
\eeq
where
\beq
\label{6.7}
\theta_0=\frac{m_0T}{m T_0}, \quad \theta_{0g}=\frac{m_0T_g}{m_g T_0}.
\eeq
While the quantity $\theta_0$ is the ratio of the mean square velocities of the intruders and granular gas particles, the quantity $\theta_{0g}$ gives the ratio of the mean square velocities of the intruders and molecular gas particles. Moreover, the explicit form of the derivative $\partial \tau_0/\partial \chi$ appearing in Eq.\ \eqref{6.3} is given in the supplementary material

In the Brownian limiting case ($m\gg m_g$ and $m_0\gg m_g$), $\widetilde{\nu}_D\to 1$, $x\to \chi^{-1/2}$, $\beta\to 1-3\chi^{-1}$ and Eq.\ \eqref{6.3} yields
\beq
\label{6.8}
D_T^*\to D_T^{*\text{B}}=\frac{\tau_0-\frac{m_0}{m}+\chi\frac{\partial \tau_0}{\partial \chi}}{\nu_D^*+\gamma_0^*-2\gamma^*\chi^{-1}-\frac{1}{2}\zeta^*}.
\eeq
Upon obtaining Eq.\ \eqref{6.8} use has been made of the steady-state condition
\beq
\label{6.9}
\chi^{-1}\gamma^*=\gamma^*+\frac{1}{2}\zeta^*.
\eeq
The expression \eqref{6.8} agrees with previous results derived from the Langevin-like model based on the Fokker--Planck operators \eqref{1.14} and \eqref{2.9} \cite[]{GG22a,GG23}.

\subsection{Mutual diffusion coefficient $D$}

The mutual diffusion coefficient $D$ is
\beq
\label{6.10}
D=\frac{n_0T}{m_0\nu}D^*, \quad D^*=\frac{\zeta^* D_T^*-\frac{m_0}{m}+\phi\frac{\partial \tau_0}{\partial \phi}}{\nu_D^*+\widetilde{\nu}_D\gamma_0^*}.
\eeq
In Eqs.\ \eqref{6.3} and \eqref{6.10}, the derivatives $\partial\tau_0/\partial\chi$ and $\partial\tau_0/\partial\phi$  can be seen as a measure of the departure of the perturbed time-dependent state from the steady reference state. Their explicit forms and derivations are provided in the supplementary material.

In the Brownian limit ($m\gg m_g$ and $m_0\gg m_g$), $\widetilde{\nu}_D\to 1$ and Eq.\ \eqref{6.10} reduces to
\beq
\label{6.12}
D^*\to D^{*\text{B}}=\frac{\zeta^* D_T^*-\frac{m_0}{m}+\phi\frac{\partial \tau_0}{\partial \phi}}{\nu_D^*+\gamma_0^*}.
\eeq
Equation \eqref{6.12} is consistent with the results obtained from the Langevin-like model \cite[]{GG22a,GG23}.

\subsection{Tracer diffusion coefficient $D_0$}

The tracer diffusion coefficient $D_0$ is given by
\beq
\label{5.15.1}
D_0=\frac{m n T}{m_0^2 \nu}D_0^*, \quad D_0^*=\frac{\tau_0}{\nu_D^*+\widetilde{\nu}_D\gamma_0^*}.
\eeq
In the Brownian limit,
\beq
\label{5.15.3}
D_0^*\to D_0^{\text{B}*}=\frac{\tau_0}{\nu_D^*+\gamma_0^*},
\eeq
which agrees with previous results \cite[]{GG22a,GG23}.

\subsection{Velocity diffusion coefficient $D_0^U$}

The diffusion coefficient $D_0^{U}$ is given by
\beq
\label{5.16}
D_0^{U}=m_0 n_0 D_0^{U*}, \quad
D_0^{U*}=\frac{\xi_0^*-\xi^*}{\nu_D^*+\widetilde{\nu}_D\gamma_0^*}.
\eeq
Here,
\beq
\label{5.17}
\xi_0^*=\frac{\xi_0}{\rho_0\nu}=\mu_{0g}\theta_{0g}^{-1/2}\left(1+\theta_{0g}\right)^{1/2}\gamma_0^*,
\eeq
\beq
\label{5.18}
\xi^*=\frac{\xi}{\rho \nu}=\mu\theta^{-1/2}\left(1+\theta\right)^{1/2}\gamma^*,
\eeq
where $\rho_0=m_0n_0$ and $\theta=m T_g/(m_g T)$.

In the Brownian limit, $\xi_0^*\to \gamma_0^*$, $\xi^*\to \gamma^*$, $\widetilde{\nu}_D\to 1$, and Eq.\ \eqref{5.16} reduces to
\beq
\label{5.16.1}
D_0^{U\text{B}*}=\frac{\gamma_0^*-\gamma^*}{\nu_D^*+\gamma_0^*},
\eeq
which is consistent with the previous results \cite[]{GG22a,GG23} derived from the Langevin-like model.

\subsection{Self-diffusion limiting case}

Another interesting limiting case corresponds to the self-diffusion problem, namely,
when the intruders move in granular gas whose particles are  mechanically equivalent to it ($\sigma=\sigma_0$, $m=m_0$, $\al=\al_0$). In this case,  $\xi^*=\xi_0^*$, $\tau_0=1$, $\partial_\chi \tau_0=\partial_\phi\tau_0=0$ and so, Eqs.\ \eqref{6.3} and \eqref{5.16} yield, respectively,  $D_T=D_0^U=0$ as expected. Moreover, $D_0^*=-D^*$ and hence the constitutive equation \eqref{0.1} for the mass flux becomes
\beq
\label{5.17.1}
\mathbf{j}_0^{(1)}=-\frac{n T}{\nu}D_{0,\text{self}}^* \nabla x_0,
\eeq
where $x_0=n_0/n$ is the concentration (or mole fraction) of the tracer species and the self-diffusion coefficient $D_{0,\text{self}}^*$ is
\beq
\label{5.18.1}
D_{0,\text{self}}^*=\frac{1}{\nu_{D,\text{self}}^*+\gamma^*\widetilde{\nu}_{D,\text{self}}}.
\eeq
In Eq.\ \eqref{5.18},
\beq
\label{5.19}
\nu_{D,\text{self}}^*=\frac{\sqrt{2}\pi^{(d-1)/2}}{d\Gamma\left(\frac{d}{2}\right)}(1+\al),\quad \widetilde{\nu}_{D,\text{self}}=\frac{m}{m+m_g}\left(1+\frac{m_g T}{m T_g}\right)^{1/2}.
\eeq

\begin{figure}
\begin{center}
\includegraphics[width=0.55\textwidth]{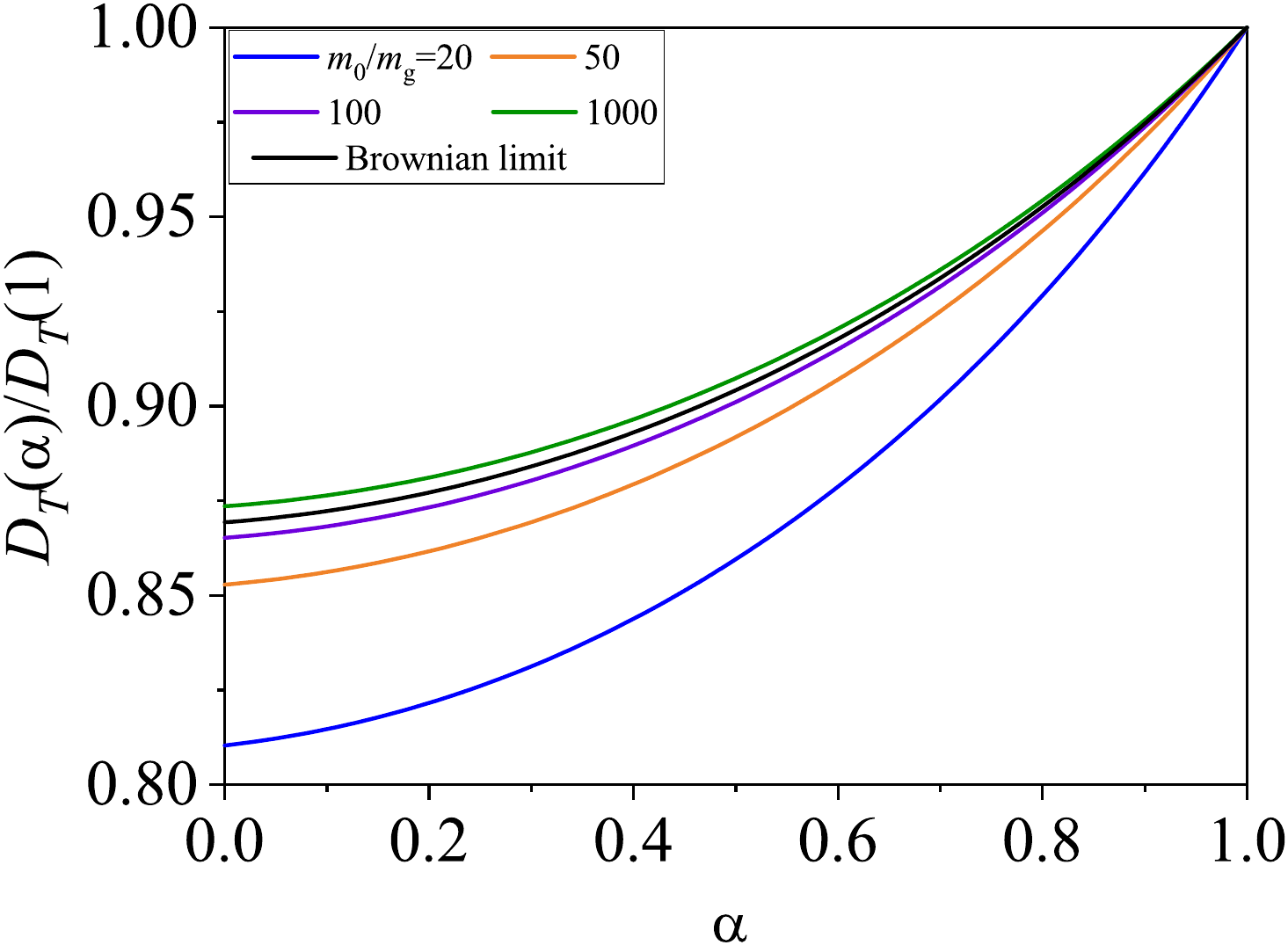}
\caption{Plot of the (scaled) thermal diffusion coefficient $D_T(\al)/D_T(1)$ versus the (common) coefficient of restitution $\al=\al_0$ for $d=3$, $\phi=0.0052$, $T_g^*=10$, and four different values of the mass ratio $m_0/m_g$ [$m_0/m_g=20, 50, 100$, and 1000]. In all the curves $m_0/m=8$, $\sigma_0/\sigma=2$,  and $\sigma_0/\sigma_g=(m_0/m_g)^{1/3}$. The dashed line refers to the expression \eqref{6.8} derived in the Brownian limiting case for the ratio $D_T(\al)/D_T(1)$.}
\label{fig5}
\end{center}
\end{figure}
\begin{figure}
\begin{center}
\includegraphics[width=0.55\textwidth]{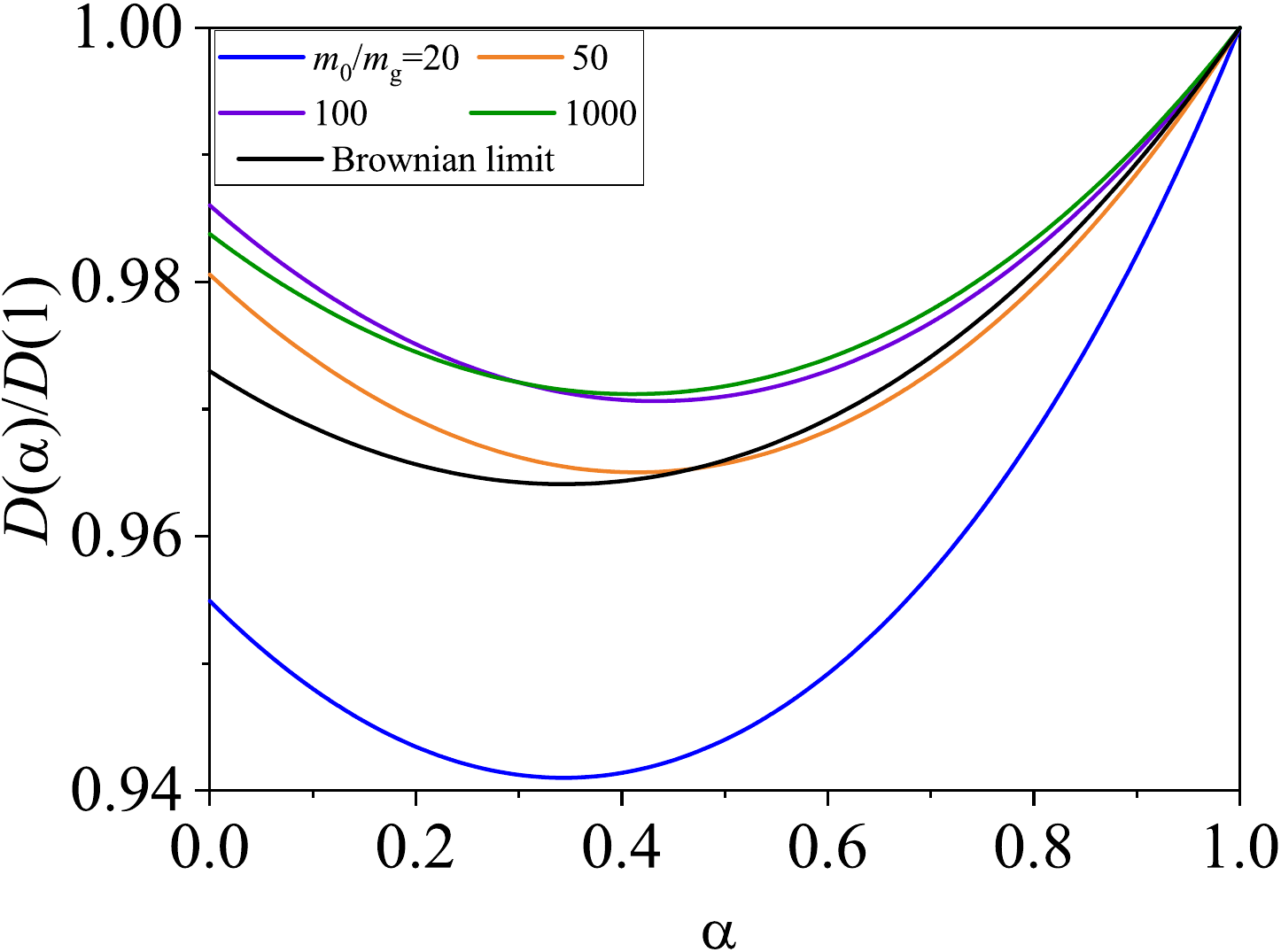}
\caption{Plot of the (scaled) mutual diffusion coefficient $D(\al)/D(1)$ versus the (common) coefficient of restitution $\al=\al_0$ for $d=3$, $\phi=0.0052$, $T_g^*=10$, and four different values of the mass ratio $m_0/m_g$ [$m_0/m_g=20, 50, 100$, and 1000]. In all the curves $m_0/m=8$, $\sigma_0/\sigma=2$, and $\sigma_0/\sigma_g=(m_0/m_g)^{1/3}$. The dashed line refers to the expression \eqref{6.12} derived in the Brownian limiting case for the ratio $D(\al)/D(1)$.}
\label{fig6}
\end{center}
\end{figure}
\begin{figure}
\begin{center}
\includegraphics[width=0.55\textwidth]{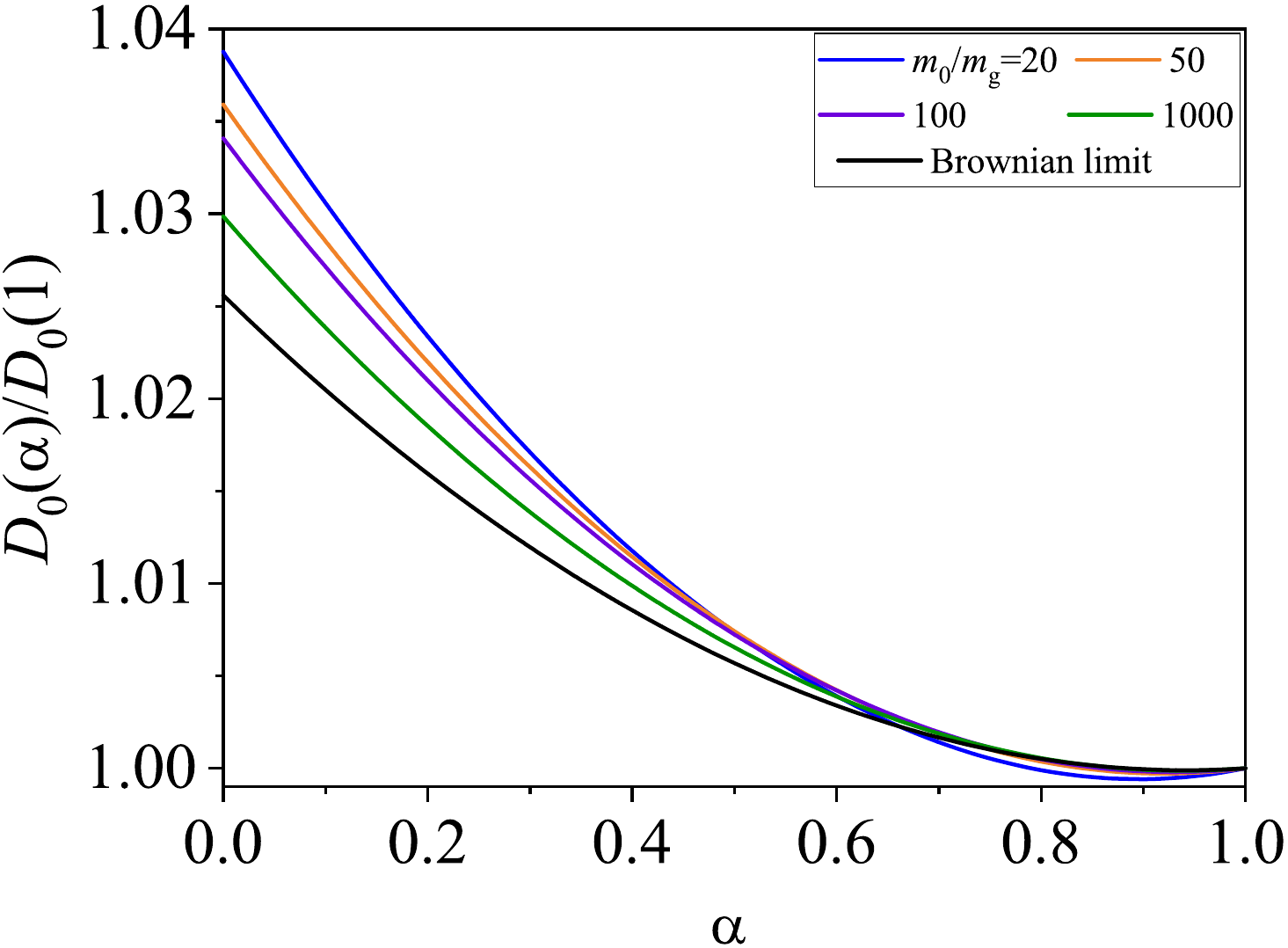}
\caption{Plot of the (scaled) tracer diffusion coefficient $D_0(\al)/D_0(1)$ versus the (common) coefficient of restitution $\al=\al_0$ for $d=3$, $\phi=0.0052$, $T_g^*=10$, and four different values of the mass ratio $m_0/m_g$ [$m_0/m_g=20, 50, 100$, and 1000]. In all the curves $m_0/m=8$, $\sigma_0/\sigma=2$, and $\sigma_0/\sigma_g=(m_0/m_g)^{1/3}$. The dashed line refers to the expression \eqref{5.15.3} derived in the Brownian limiting case for the ratio $D_0(\al)/D_0(1)$.}
\label{fig7}
\end{center}
\end{figure}
\begin{figure}
\begin{center}
\includegraphics[width=0.55\textwidth]{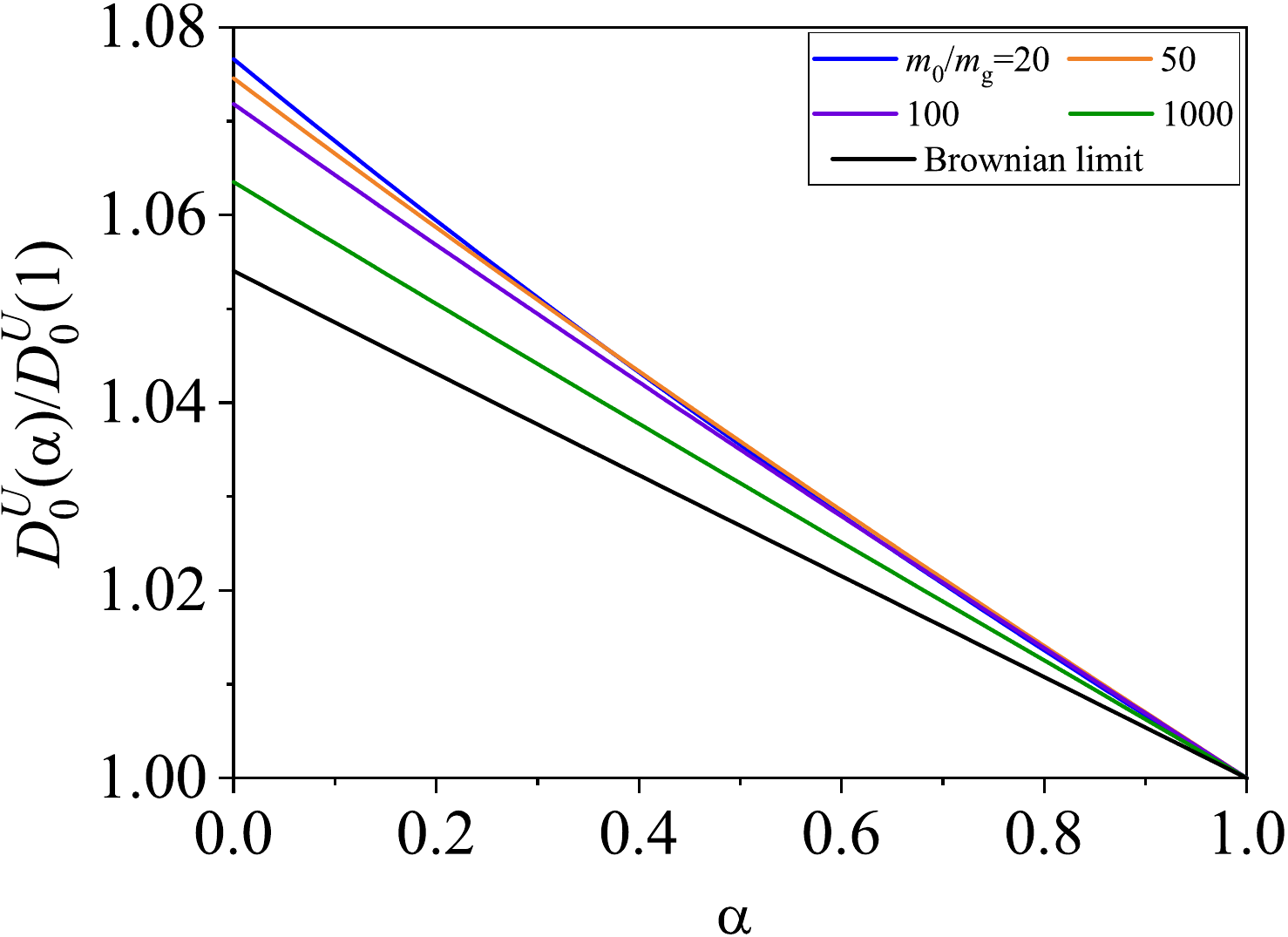}
\caption{Plot of the (scaled) velocity diffusion coefficient $D_0^{U}(\al)/D_0^{U}(1)$ versus the (common) coefficient of restitution $\al=\al_0$ for $d=3$, $\phi=0.0052$, $T_g^*=10$, and four different values of the mass ratio $m_0/m_g$ [$m_0/m_g=20, 50, 100$, and 1000]. In all the curves $m_0/m=8$, $\sigma_0/\sigma=2$, and $\sigma_0/\sigma_g=(m_0/m_g)^{1/3}$. The dashed line refers to the expression \eqref{5.16.1} derived in the Brownian limiting case for the ratio $D_0^{U}(\al)/D_0^{U}(1)$.}
\label{fig8}
\end{center}
\end{figure}

\subsection{Some illustrative systems}

The expressions of the diffusion transport coefficients $D_T$, $D$, $D_0$ and $D_{0}^U$ in the steady state are given by Eqs.\ \eqref{6.3}, \eqref{6.10}, \eqref{5.15.1}, and \eqref{5.16}, respectively. As usual in the study of transport properties in dry granular gases \cite[]{BDKS98,GD99a}, to assess the impact of inelasticity in collisions on diffusion, the diffusion transport coefficients are scaled with respect to their values for elastic collisions ($\al=\al_0=1$). As expected, these scaled diffusion coefficients depend in a complex way on the parameter space [defined by the set \eqref{3.18}] of the system. Since the parameter space $\xi$ is large, for the sake of simplicity, henceforth we consider hard spheres ($d=3$) with $\phi=0.0052$ (very dilute granular gas), $T_g^*=10$ and with a common coefficient of restitution $\al=\al_0$

Figures \eqref{fig5}--\eqref{fig8} show the $\al$-dependence of the (scaled) coefficients $D_T(\al)/D_T(1)$,  $D(\al)/D(1)$,  $D_0(\al)/D_0(1)$, and $D_0^{U}(\al)/D_0^{U}(1)$, respectively. Four different values of the mass ratio $m/m_g$ were considered; in all systems $m_0/m=8$ and $\sigma_0/\sigma=2$ [$\sigma_0/\sigma=(m_0/m_g)^{1/3}$]. In addition, the results derived from the Brownian limiting case are also shown for comparison. In this case, the diffusion coefficients $D_T$, $D$, $D_0$, and $D_0^U$ are given by Eqs.\ \eqref{6.8}, \eqref{6.12}, \eqref{5.15.3}, and \eqref{5.16.1}, respectively. \vicente{The main objective of figures \eqref{fig5}--\eqref{fig8} is to show how the combined effect of both inelasticity and the mass ratio $m_0/m_g$ affects the diffusion of intruders in a granular suspension. Additionally, since our results apply to arbitrary values of the mass ratio, we aim to evaluate for practical purposes the conditions under which the diffusion coefficients converge to their values in the so-called Brownian limit case.}

We observe that in general the diffusion transport coefficients deviate from their elastic forms, especially for strong inelasticity as expected. However, these deviations are much smaller than those have been found for \emph{dry} (no gas phase) granular mixtures (see for example, figures 4, 5 and 6 of \cite{GMV13a} for $x_0=0.2$). This means that the surrounding molecular gas generally hinders the diffusion of tracer particles in a granular gas. While the (scaled) thermal diffusion and tracer diffusion coefficients exhibit a monotonic dependence on the coefficient of restitution [$D_T(\al)/D_T(1)$ ($D_0(\al)/D_0(1)$) decreases (increases) with inelasticity] regardless the mass ratio considered, the (scaled) mutual diffusion coefficient $D(\al)/D(1)$ has a non-monotonic dependence on $\al$. \vicente{This kind of trend is not entirely new: a similar behaviour has already been analysed using a random-walk interpretation in the Fokker--Planck model \cite[]{GABG23}, where the effect was attributed to the competition between two opposite trends: (i) the decrease of the effective mean free path with increasing $\alpha$, which reduces the persistence of the intruders' trajectories, and (ii) the increase of the collision frequency in the quasielastic regime, which enhances the number of effective steps. The balance between these two competing tendencies explains the emergence of non-monotonicities in the tracer diffusion coefficient. In the present collisional model, the same physical mechanism could explain the observed results. } 

With respect to the mass ratio, for a fixed value of the (common) coefficient of restitution, it is quite obvious that the (scaled) thermal diffusion coefficient increases with increasing the mass ratio, while the (scaled) tracer diffusion coefficient decreases with increasing the mass ratio. The behaviour of the (scaled) mutual diffusion coefficient depends on the range of values of $\al$ considered since there are crossings between the different curves. The (scaled) velocity diffusion coefficient has no analogue in the dry granular case. We see that it always increases with increasing inelasticity. Moreover, at a given value of $\al$, it increases with decreasing the mass ratio $m/m_g$. We also observe that the convergence of the curves towards the Brownian limiting case is slower than that found by \cite[]{GG22} for \emph{monocomponent} granular suspensions. In this latter case, the results derived for finite values of $m/m_g$ practically coincide with those obtained in the Brownian limit for values of the mass ratio around 50. Figures \eqref{fig5}--\eqref{fig8} clearly show that there are still (small) discrepancies between the results derived here and those obtained in the Brownian limit  for values of the mass ratio $m_0/m_g=1000$.

\section{An application: Segregation of intruders in a granular suspension}
\label{sec7}

As mentioned in section \ref{sec1}, an interesting application of the results displayed in section
\ref{sec6} is the the study of the segregation of intruders by thermal diffusion in a granular suspension. Needless to say, thermal diffusion segregation is likely one of the most common phenomenon appearing in polydisperse systems. It occurs in a non-convective steady state ($\mathbf{U}=\mathbf{U}_g=\mathbf{0}$) due to the existence of a temperature gradient, which causes the movement of the different species of the mixture. In the steady state, remixing of species generated by diffusion is balanced by segregation caused by temperature differences. To quantify the degree of segregation along the temperature gradient is usual to introduce the thermal diffusion factor $\Lambda$ \cite[]{KCL87}.
In a steady state without convection and where the mass flux is zero ($\mathbf{j}_0^{(1)}=\mathbf{0}$), the thermal diffusion factor is defined as
\beq
\label{7.1}
-\Lambda \frac{\partial \ln T}{\partial z}=\frac{\partial}{\partial z}\ln \Big(\frac{n_0}{n}\Big),
\eeq
where for the sake of simplicity we have assumed that gradients occur only along the axis $z$. In addition, we also assume that the gravitational field is parallel to the thermal gradient, namely, $\mathbf{g}=-g\widehat{e}_z$, where $\widehat{e}_z$ is the unit vector in the positive direction of the $z$ axis.

\begin{figure}
\begin{center}
\includegraphics[width=0.55\textwidth]{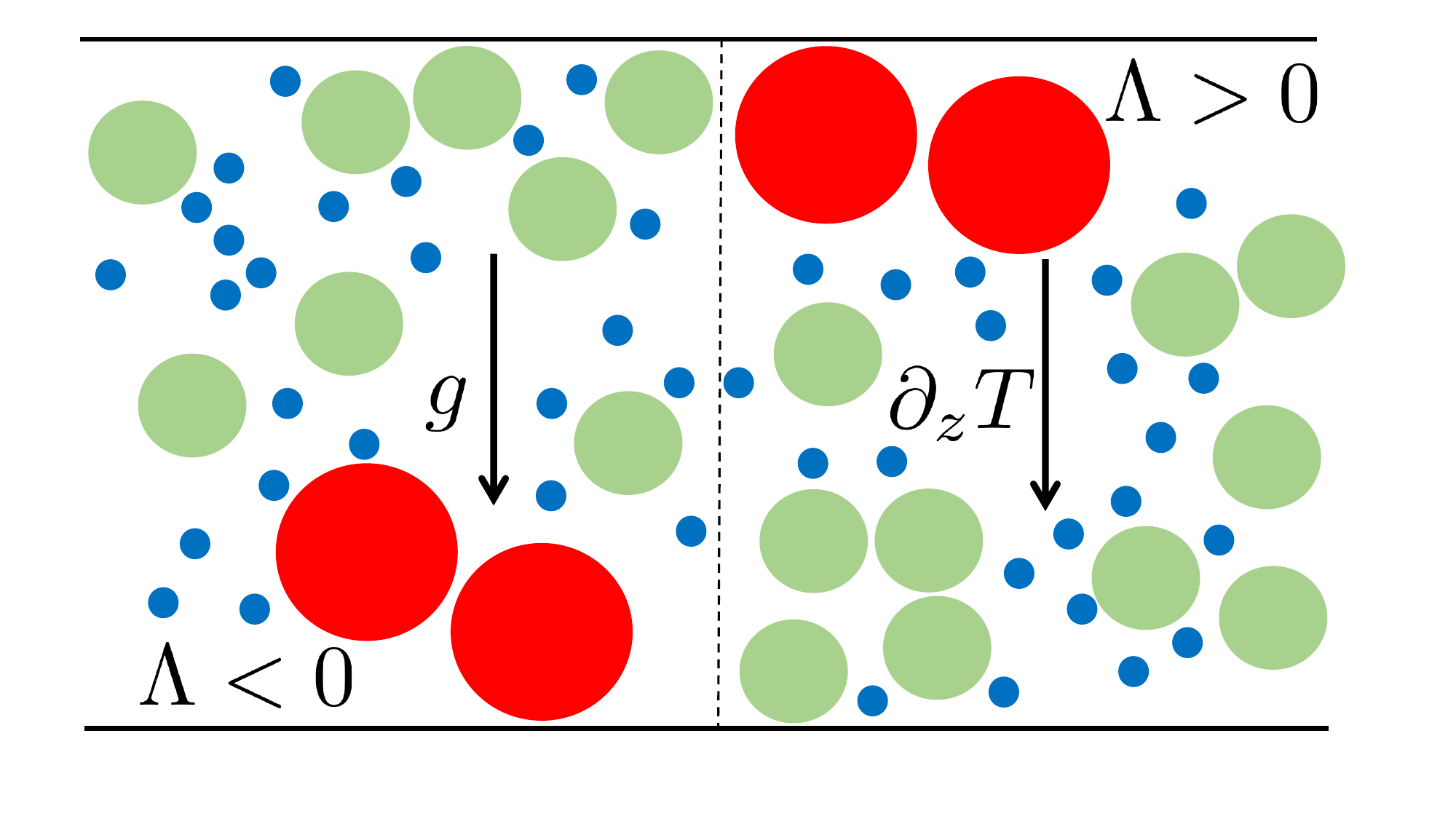}
\caption{A representative diagram of the BNE ($\Lambda>0$) and RBNE ($\Lambda<0$) effects for a ternary system composed of molecular particles (blue), granular particles (green), and intruders (red).}
\label{figdiag}
\end{center}
\end{figure}

We consider a scenario in which the intruders have a larger size than the granular gas particles ($\sigma_0>\sigma$). Furthermore, as said before, since gravity and the thermal gradient point in the same direction, then the lower plate is hotter than the upper plate ($\partial_z\ln T<0$). Based on Eq.\ \eqref{7.1}, when $\Lambda>0$, intruders rise relative to granular gas particles ($\partial_z \ln (n_0/n)>0$), leading to an accumulation of tracer particles near the cooler plate. This situation is commonly known as the Brazil nut effect (BNE). Conversely, for $\Lambda<0$, intruders fall relative to granular gas particles ($\partial_z \ln (n_0/n)<0$), resulting in an accumulation near the hotter plate. This situation is known as the Reverse Brazil Nut Effect (RBNE). A representative diagram of segregation dynamics is shown in figure \ref{figdiag}.

We express the thermal diffusion factor in terms of the diffusion transport coefficients. In the steady state, to first order in spatial gradients, the momentum balance \eqref{1.9} reduces to
\beq
\label{7.2}
\left(1-\varepsilon \kappa^*\right)+\left(1-\varepsilon \overline{\mu}^*\right)\frac{\partial_z \ln n}{\partial_z \ln T}=-g^*,
\eeq
where
\beq
\label{7.3}
g^*=\frac{\rho g}{n\partial_z T}<0
\eeq
is a dimensionless parameter measuring the gravity relative to the thermal gradient, $\kappa^*$ and $\overline{\mu}^*$ are the (reduced) thermal conductivity and diffusive heat conductivity transport coefficients, respectively, and
\beq
\label{7.4}
\varepsilon=\frac{(d+2)\sqrt{\pi}}{2^{d+3}(d-1)}\frac{\mu \chi^{-1/2}}{\phi \sqrt{T_g^*}}X(\theta).
\eeq
Here, $X(\theta)=\theta^{-1/2}(1+\theta)^{-1/2}$. The explicit forms of $\kappa^*$ and $\overline{\mu}^*$ are displayed in the supplementary material. Upon obtaining Eq.\ \eqref{7.2}, use has been made of the result in the leading Sonine approximation \cite[]{GG22}
\beq
\label{7.5}
\mathcal{F}^{(1)}[f^{(1)}]\to \frac{1}{d+2}\frac{\rho \mu \gamma}{n}\kappa_0 X(\theta) \Big(\kappa^* \partial_z \ln T+\overline{\mu}^* \partial_z \ln n\Big),
\eeq
where
\beq
\label{7.5.1}
\kappa_0=\frac{d(d+2)^2\Gamma\left(\frac{d}{2}\right)}{16(d-1)\pi^{\frac{d-1}{2}}}\sigma^{1-d}\sqrt{\frac{T}{m}}
\eeq
is the low-density value of the thermal conductivity for an ordinary gas of hard spheres.

According to Eq.\ \eqref{0.1}, when $\Delta \mathbf{U}=\mathbf{0}$, the condition $j_{0,z}^{(1)}=0$ yields the relation
\beq
\label{7.6}
-D_0^* \partial_z \ln n_0=D^* \partial_z \ln n+D_T^* \partial_z \ln T,
\eeq
where $D_T^*$, $D^*$, and $D_0^*$ are given by Eqs.\ \eqref{6.3}, \eqref{6.10}, and \eqref{5.15.1}, respectively. From Eqs.\ \eqref{7.2} and \eqref{7.6}, the thermal diffusion factor can be written as
\beq
\label{7.7}
\Lambda=\frac{\partial_z \ln n}{\partial_z \ln T}-\frac{\partial_z \ln n_0}{\partial_z \ln T} =\frac{D_T^*+\left(D_0^*+D^*\right)\left(\varepsilon\kappa^*-1-g^*\right)\left(1-\varepsilon\overline{\mu}^*\right)^{-1}}{D_0^*}.
\eeq

In the Brownian limit ($m/m_g\to \infty$), $\mu \to 1$, $\theta\to \infty$, so that $\varepsilon \to 0$. In this limiting case, $\Lambda$ reduces to 
\beq
\label{7.7.1}
\Lambda=\frac{D_T^*-\left(D_0^*+D^*\right)\left(1+g^*\right)
}{D_0^*}.
\eeq
Equation \eqref{7.7.1} agrees with previous results derived for the segregation of massive intruders in a granular suspension \cite[]{GG22a}.

Since Eq.\ \eqref{5.15.1} clearly shows that $D_0^*>0$, then the curves delineating the regimes between the segregation toward the cold and the hot wall (BNE/RBNE transition) are determined from the condition
\beq
\label{7.8}
\left(1-\varepsilon\overline{\mu}^*\right)D_T^*=-\left(D_0^*+D^*\right)\left(\varepsilon\kappa^*-1-g^*\right).
\eeq

\subsection{Mechanically equivalent particles}

In this scenario, $D_T^*=0$ and $D^*=-D_0^*$, thus Eq.\ \eqref{7.8} is valid for any values of the coefficients of restitution, masses, and diameters. Consequently, as in the Brownian limit \cite[]{GG23}, no segregation occurs in the system.

\subsection{Different mechanical properties}
\subsubsection{Absence of gravity ($|{g}^*|\to 0$)}

Let us first consider a scenario where gravitational effects are negligible ($|{g}^*|\to 0$). Under this assumption, the condition $\Lambda=0$ reads
\beq\label{7.9}
\left(1-\varepsilon\overline{\mu}^*\right)D_T^*=-\left(D_0^*+D^*\right)\left(\varepsilon\kappa^*-1\right).
\eeq

\begin{figure}
\begin{center}
\includegraphics[width=0.55\textwidth]{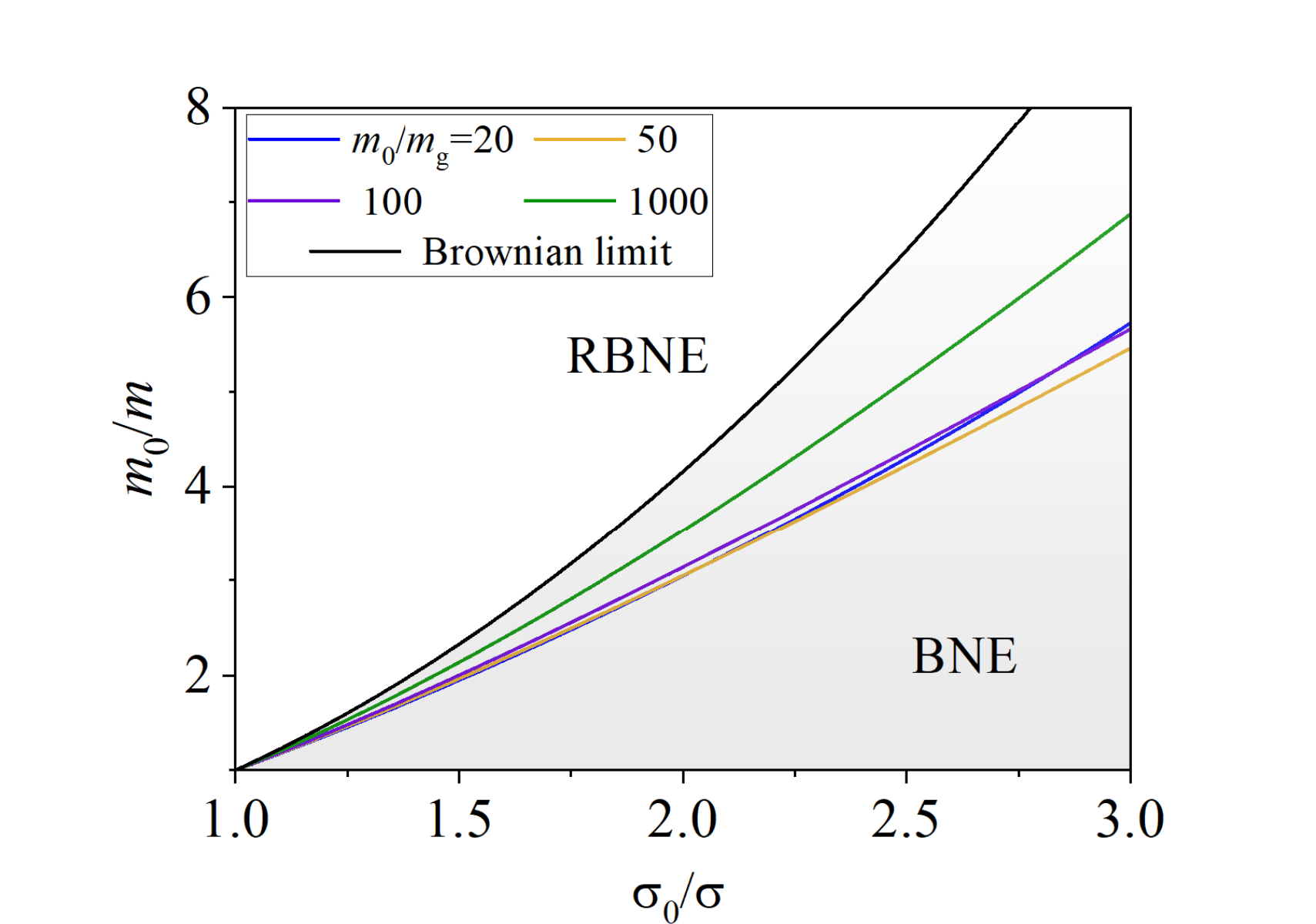}
\caption{Plot of the marginal segregation curve ($\Lambda=0$) for a (common) coefficient of restitution $\al=\al_0=0.7$ and $|{g}^*|\to 0$. The parameters used are: $d=3$, $\phi=0.0052$, $T_g^*=10$, and four different values of the mass ratio $m_0/m_g$ [$m_0/m_g$=20, 50, 100, and 1000]. The points below the curve correspond to $\Lambda>0$ (BNE), while the points above the curve correspond to $\Lambda<0$ (RBNE). The dashed line is the result obtained in the Brownian limiting case.}
\label{figseg1}
\end{center}
\end{figure}

Figure \ref{figseg1} shows the BNE/RBNE phase diagram for a three-dimensional system ($d=3$) with a common coefficient of restitution $\alpha=\alpha_0=0.7$, $T_g^*=10$, and five distinct values of the mass ratio $m_0/m_g$ ($m_0/m_g=20, 50, 100, 1000$, and the Brownian limit $m_0/m_g\to \infty$). At first glance, we notice a quantitative discrepancy with the diagrams shown in \cite{GG23} for $|{g}^*|\to 0$. These discrepancies were expected since, in that work, segregation was calculated for moderate densities, unlike here ($\phi=0.0052$). Also, when using the Fokker--Planck approach \eqref{2.9} for the operator $J_{0g}[f_0,f_g]$, we see that the expression \eqref{3.21.1} of $\gamma_0^*$ employed here slightly differs from the expression of $\gamma^*_{0,\text{FP}}$ defined by \cite{GG23}. Specifically, $\gamma^*_{0,\text{FP}}=(\sigma_0/\sigma)(\overline{\sigma}/\overline{\sigma}_0)^{d-1}\gamma_0^*$. This discrepancy is because $\gamma^*_{0,\text{FP}}$ was obtained from the granular literature using phenomenological arguments. 

However, what is surprising is the complete reversal we see in the diagram when we observe the RBNE effect as we increase $m_0/m$, contrary to what is observed in  \cite{GG23}. This requires a more subtle explanation. When calculating the diagrams in the case of an effective model where the thermostat is modelled by a Fokker--Planck equation, the thermostat intensity is regulated only by the (dimensionless) bath temperature $T_{g}^*$. If this parameter is kept constant, as in the figures 11 and 12 reported by  \cite{GG23}, the thermostat does not change when the mass ($m_0/m$) or size ($\sigma_0/\sigma$) ratios are modified. However, in our case, we modify $m_0/m$ (or $\sigma_0/\sigma$) keeping $m_0/m_g$ constant for each particular curve. Thus, by modifying the mass ratio $m_0/m$ for a particular value of $m_0/m_g$, the relative mass between the granular and molecular gas changes. Therefore, the effect of the collisions between the grains and the particles of the molecular gas will be different. Concretely, if we increase $m_0/m$ without changing $m_0/m_g$, the molecular gas will have a mass increasingly similar to that of the grains. Consequently, the thermalising effect of the molecular gas that compensates for the effect of inelasticity will be more effective, causing the temperature of the granular gas to be higher and thus tend to increase with respect to that of the intruders (RBNE). We observe the same when we increase the size ratio ($\sigma_0/\sigma$). In this case, for a given $m_0/m_g$ (or equivalently $\sigma_0/\sigma_g$), as the size of the intruders increases, the grains will have a size increasingly similar to the particles of the surrounding molecular gas, thus the effective area of the grain in a collision decreases, and with it, the number of collisions. Therefore, the grain will have less effective thermalization and will move to the cold zone (BNE).

On the other hand, as we increase $m_0/m_g$, we see that the transition to the RBNE phase occurs earlier. This is because, by increasing the mass more rapidly than the size [$m_0/m_g=(\sigma_0/\sigma_g)^d$], the effect of intruders-molecular gas collisions in the motion of intruders becomes less effective. In the curves where the mass ratio $m_0/m_g$ is close to each other, the relative size also plays an important role, and crossings can be observed, as was the case in figure \ref{fig6}.

Moreover, for elastic collisions ($\alpha = \alpha_0 = 1$), the segregation criterion notably deviates from the classical result obtained for molecular mixtures of hard spheres by \cite{KCL87}, where, in the first Sonine approximation, the condition $\Lambda=0$ yields the simple segregation criterion $m_0/m = 1$. For the present system (intruders moving in a granular gas immersed into a molecular gas), our analysis reveals that there is no segregation for the remaining parameters considered in figure \ref{figseg1}.

\subsubsection{Thermalized systems ($\partial_z T\to 0$)}

Let us explore a scenario where gravity is the main factor influencing segregation dynamics. In this situation, $|{g}^*|\to \infty$, which makes the temperature gradient negligible ($\partial_y T\to 0$), and the condition $\Lambda=0$ leads to the relation
\beq
\label{7.10}
D_{0}^*+D^*=0.
\eeq

\begin{figure}
\begin{center}
\includegraphics[width=0.55\textwidth]{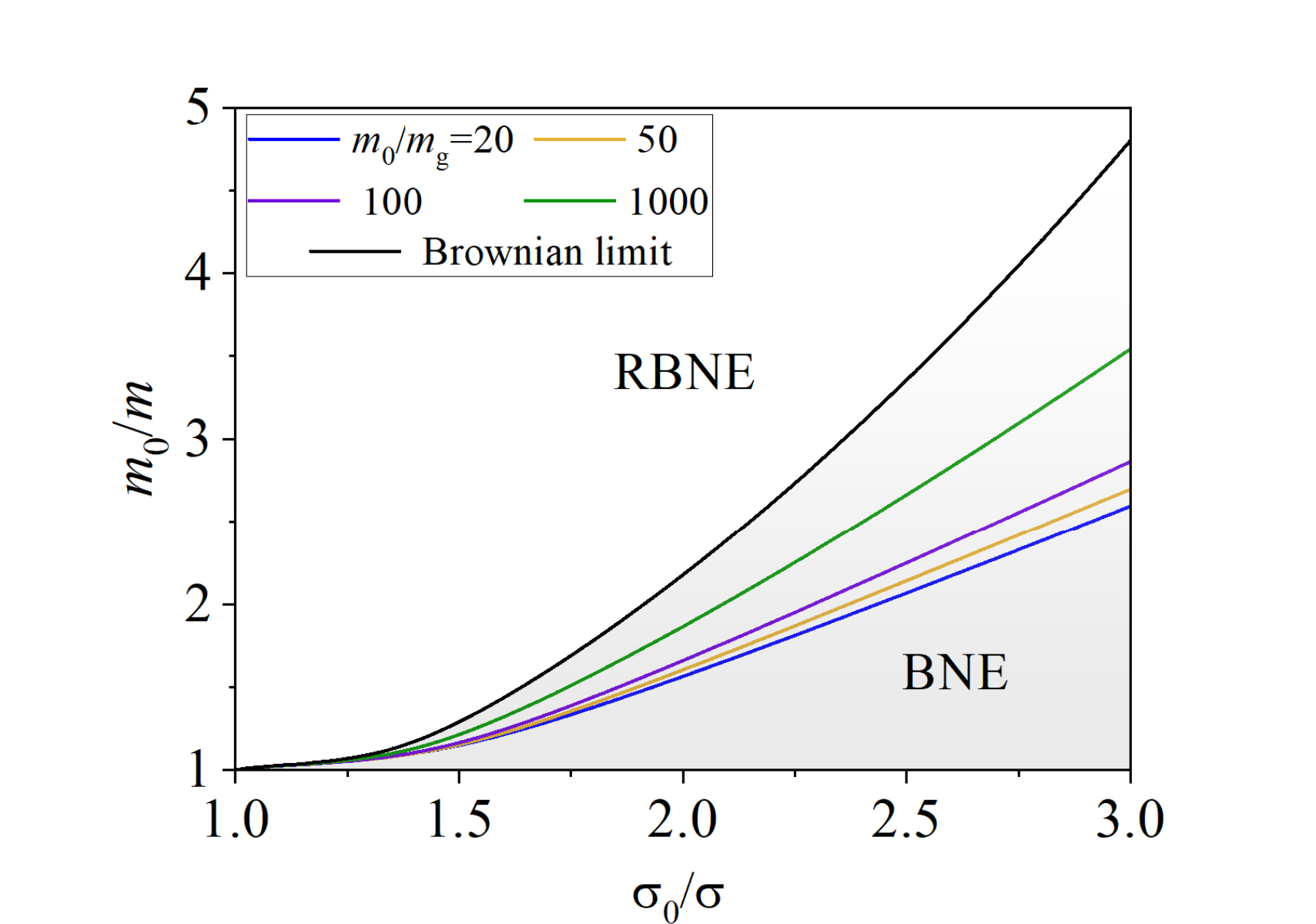}
\caption{Plot of the marginal segregation curve ($\Lambda=0$) for a (common) coefficient of restitution $\al=\al_0=0.7$ and and $|{g}^*|\to \infty$. The parameters used are: $d=3$, $\phi=0.0052$, $T_g^*=10$, and four different values of the mass ratio $m_0/m_g$ [$m_0/m_g=20, 50, 100$, and 1000]. The points below the curve correspond to $\Lambda>0$ (BNE), while the points above the curve correspond to $\Lambda<0$ (RBNE). The dashed line is the result obtained by using the Fokker--Planck approach \eqref{1.14} to the operator $J_g[f,f_g]$.}
\label{figseg2}
\end{center}
\end{figure}

\vicente{Figure \ref{figseg2} shows the case $|{g}^*|\to \infty$ with the same parameters as Fig. \ref{figseg1}. In this strong-gravity case, the explanation is simpler. Gravity is much stronger than both the energy coming from the molecular gas and the energy lost in inelastic collisions, as seen in granular suspensions and driven granular gases \cite[]{G08a,GG23,GGBS24}. Because of this, all particles fall to the bottom plate. Heavier intruders are harder to lift, so they move downward (RBNE). However, if the intruders are bigger but keep the same mass, smaller particles hit it more often. These collisions push the smaller particles below the intruders, lifting them and creating a buoyancy effect (BNE)}

For elastic collisions ($\al=\al_0=1$), the cooling rate vanishes ($\zeta^*=0$), and there is equipartition of energy ($\tau_0=1$).Thus, it is straightforward to verify from Eqs.\ \eqref{6.10} and \eqref{5.15.1} that the segregation criterion simplifies to 
$m_0/m=1$. This result aligns with previous findings for dry granular gases (i.e., in the absence of an interstitial gas phase) \cite[]{BRM05,G06,G11}, as well as with those obtained in the coarse-grained approach by using the Fokker–Planck equation \cite[]{GG23}.

\subsubsection{General case}

\begin{figure}
\begin{center}
\includegraphics[width=0.55\textwidth]{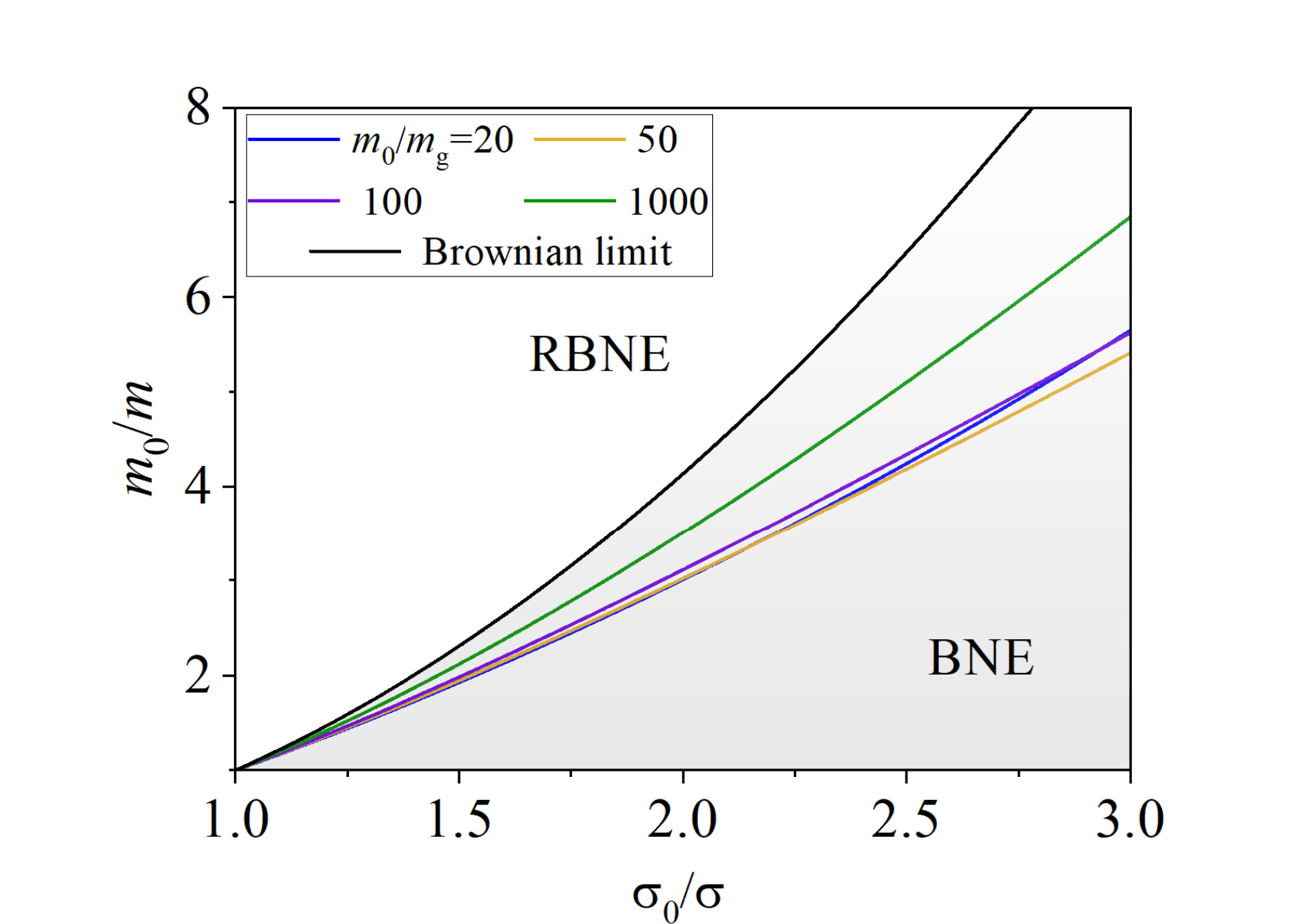}
\caption{Plot of the marginal segregation curve ($\Lambda=0$) for a (common) coefficient of restitution $\al=\al_0=0.7$ and and $|{g}^*|=1$. The parameters used are: $d=3$, $\phi=0.0052$, $T_g^*=10$, and four different values of the mass ratio $m_0/m_g$ [$m_0/m_g=20, 50, 100$, and 1000]. The points below the curve correspond to $\Lambda>0$ (BNE), while the points above the curve correspond to $\Lambda<0$ (RBNE). The dashed line is the result obtained by using the Fokker--Planck approach \eqref{1.14} to the operator $J_g[f,f_g]$.}
\label{figseg3}
\end{center}
\end{figure}

As a final situation, we consider the general case where the effects of the temperature gradient and gravity are comparable. To exemplify this, Fig. \eqref{figseg3} shows the marginal segregation curve for a reduced gravity $|{g}^*|=1$ and for the same systems depicted in figures \ref{figseg1} and \ref{figseg2}.

\begin{figure}
\begin{center}
\includegraphics[width=0.55\textwidth]{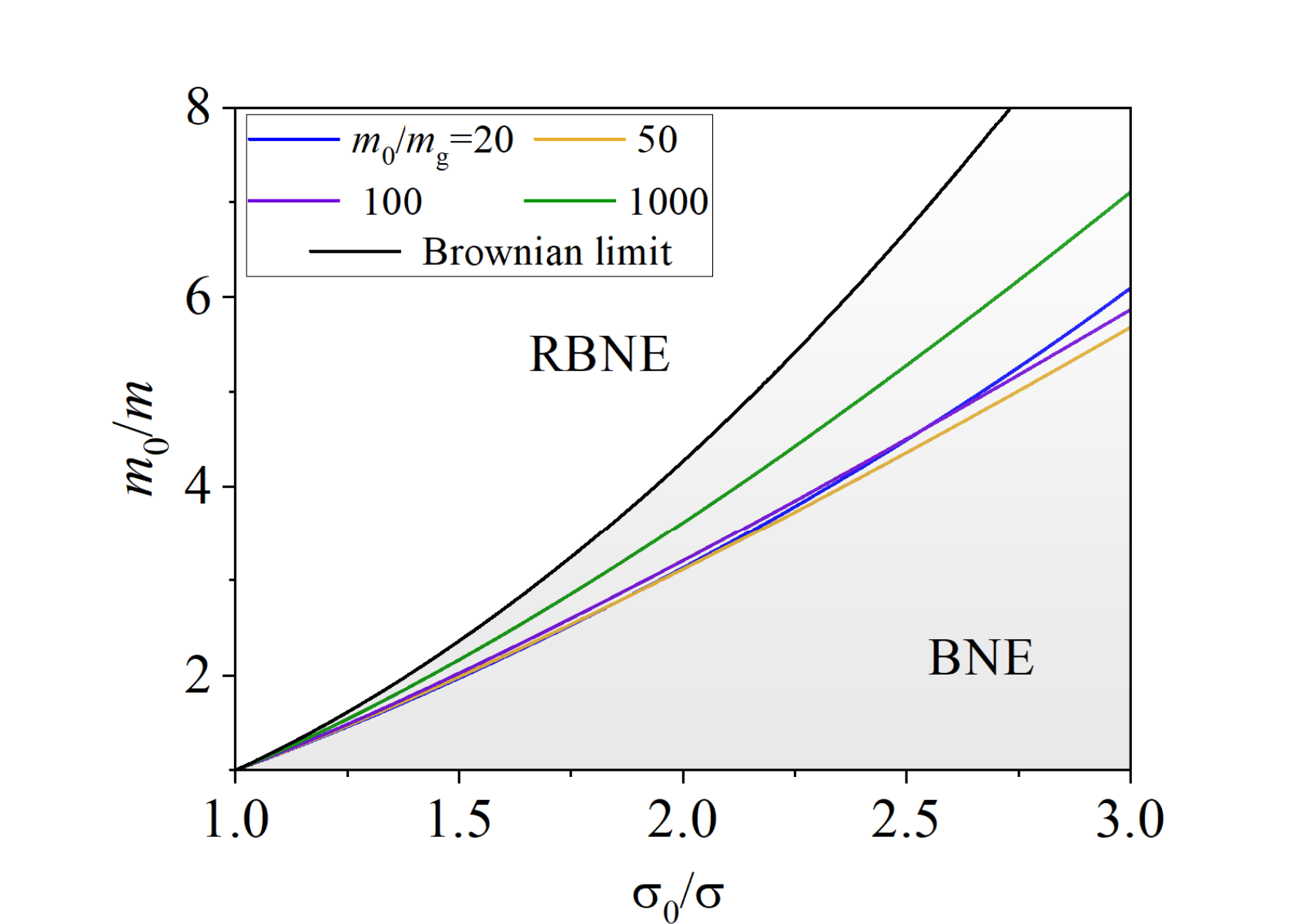}
\caption{Plot of the marginal segregation curve ($\Lambda=0$) for a (common) coefficient of restitution $\al=\al_0=1$ and and $|{g}^*|=1$. The parameters used are: $d=3$, $\phi=0.0052$, $T_g^*=10$, and four different values of the mass ratio $m_0/m_g$ [$m_0/m_g=20, 50, 100$, and 1000]. The points below the curve correspond to $\Lambda>0$ (BNE), while the points above the curve correspond to $\Lambda<0$ (RBNE). The dashed line is the result obtained by using the Fokker--Planck approach \eqref{1.14} to the operator $J_g[f,f_g]$.}
\label{figseg4}
\end{center}
\end{figure}

The main point is that, unlike in dry granular mixtures and granular suspensions \cite[]{G19,GG23,GGBS24}, gravity has a weaker effect than the thermal gradient, keeping the BNE/RBNE transition quite similar to that shown in figure \ref{figseg1}. The explanation can be the same as in the case without gravity since, when increasing $m_0/m$, the relative mass of the grains compared to molecular particles decreases. As a result, the thermalization due to the surrounding fluid does not remain constant, making thermal effects more pronounced even when gravity is present. 
To complement the phase diagram shown in figure \ref{figseg3} for inelastic collisions, in figure \ref{figseg4} we plot the marginal segregation curve ($\Lambda=0$) for the same values of the mass ratio $m_0/m_g$ as in figure \ref{figseg3} but for \textit{elastic} collisions. It is quite apparent from the comparison between figures \ref{figseg3} and \ref{figseg4} that the inelasticity of collisions plays a secondary role in the segregation behaviour in this case since we observe practically no differences between both figures.

\section{Summary and concluding remarks}
\label{sec8}

This paper aims to determine the diffusion transport coefficients for tracer (or intruder) particles within a granular gas modelled as a gas of inelastic hard spheres. Intruders and granular gas are
immersed in a bath of elastic hard spheres (molecular gas). We examine scenarios where the granular particles are sufficiently dilute, ensuring the molecular gas remains unaffected and serves as a thermostat at a given temperature $T_g$. Unlike other suspension models, which consider an effective fluid-solid force, our model accounts for both inelastic collisions between the tracer and granular particles, as well as among the granular particles themselves. Additionally, it takes into account 
the elastic collisions between the grains and molecular gas particles, as well as between the intruders and molecular gas. We consider the low-density regime for the suspended solid particles and hence,  the velocity distribution functions $f(\mathbf{r}, \mathbf{v}; t)$ and  $f_0(\mathbf{r}, \mathbf{v}; t)$ for grains and intruders, respectively, obey the (inelastic) Boltzmann equations.

To ensure a consistent theoretical framework, we first analyse homogeneous reference states as a basis for applying the Chapman--Enskog method and obtaining then the diffusion transport coefficients. In the homogeneous state, we present new results for the 
temperature ratio $\chi_0 = T_0/T_g$ and the kurtosis $c_0$ associated with the intruders. As in the case of the granular gas \cite[]{GG22}, we find that the tracer temperature ratio $\chi_0 $ shows larger deviations from unity as the mas ratio $m_0/m_g$ decreases. Regarding the kurtosis $c_0$ (which measures the departure of the tracer distribution from its Maxwellian form), as for the granular gas \cite[]{GG22}, our results clearly show that this quantity 
remains small enough to validate the use of the Maxwellian approximation for the velocity distributions.

We compared the theoretical results derived in the HSS for $\chi_0$ and $c_0$ with DSMC data, and the agreement is remarkable. This justifies the use of the Maxwellian distribution \eqref{3.17} to achieve accurate estimates of the cooling rates $\zeta_0$ and $\zeta_{0g}$. We also observe the convergence of the results in the Brownian limit ($m/m_g \to \infty$ and $m_0/m_g \to \infty$) with those obtained in previous works \cite[]{GGG19a,GG23} by using the Fokker--Planck approach.

Once the homogeneous state is characterized, the corresponding set of kinetic equations for the mixture were addressed using the Chapman--Enskog method \cite[]{CC70}, up to the first order in spatial gradients. From this solution, the four diffusion transport coefficients are derived by considering the leading terms in a Sonine polynomial expansion of the first-order distribution function. In dimensionless form, these diffusion coefficients are given in terms of eight dimensionless parameters: the diameter ratios $\sigma/\sigma_g$ and $\sigma_0/\sigma_g$, the mass ratios $m/m_g$ and $m_0/m_g$, the coefficients of restitution $\al$ and $\al_0$, the (reduced) bath temperature $T_g^*$, and the volume fraction $\phi$ (which is considered to be quite small since we are considering  a very dilute granular gas). Compared to previous attempts to obtain tracer diffusion coefficients in granular suspensions (see, for example, \cite[]{GG23}) by starting from a coarse-grained approach, our results provide a more general description of the dependence of diffusion on mass ($m/m_g$ and $m_0/m_g$) and diameter ($\sigma/\sigma_g$ and $\sigma_0/\sigma_g$) ratios.  As expected, the expressions for the tracer diffusion coefficients agree with previous results derived in the Brownian limit from the Fokker–Planck operator \cite[]{GG23}. Thus, with respect to previous attempts for obtaining tracer diffusion coefficients in granular suspensions \cite[]{GG23} by starting from a coarse-grained approach, our results provide a more general description about the dependence of the diffusion on the mass ($m/m_g$ and $m_0/m_g$) and diameter ($\sigma/\sigma_g$ and $\sigma_0/\sigma_g$) ratios. In particular, as expected, the expressions for the tracer diffusion coefficients agree with previous results derived in the Brownian limit from the Fokker--Planck operator  \cite[]{GG23}. In this context, the present results encompass the previous ones as a special case, thus covering interesting physical situations that had not previously been analysed from a theoretical perspective. 

In general, the diffusion coefficients exhibit significant deviations from their elastic counterparts, particularly under strong inelasticity. A key result is that convergence to the predictions of the effective model is achieved only at large mass ratios $m_0/m_g \approx 1000$, which contrasts with the convergence observed in a monocomponent granular suspension where agreement was found at $m/m_g= 50$ \cite[]{GG22a}. This emphasizes the ability of the model to capture more realistic scenarios beyond the scope of Fokker--Planck-based models.

\vicente{As an application, we investigate segregation driven by both thermal gradients and gravity. We find that increasing the mass ratio $m_0/m_g$ tends to push the intruders toward the bottom plate (RBNE), regardless of the gravitational strength. This behaviour contrasts with the results derived from the Fokker--Planck model \cite[]{GG23}. In that model, the thermostat effect remains constant (set by $T_g^*$), whereas in our model it depends on $m_0/m_g$, resulting in different thermalisation dynamics. The main novelty of the study reported here, compared with previous works on dry granular mixtures [see e.g., \cite{G08a,G11}], is the analysis of the role of $m_0/m_g$ at fixed $\al$. Unlike the dry case, where inelasticity strongly affects BNE/RBNE transitions due to energy non-equipartition, here its impact is reduced because the molecular gas injects energy homogeneously, compensating collisional dissipation and effectively suppressing partial temperature differences through thermalisation.}

\vicente{The results reported in this work suggest a framework to study experimentally the influence of an interstitial gas on segregation dynamics. In this context, a feasible experiment can be designed following the approach discussed in \cite{GABG23}. A suitable system to represent the low-Reynolds-number regime, in which the effect of granular collisions is comparable to the effect of the thermal bath, can be realized by immersing gold grains in hydrogen gas at low pressure. In this setup, the molecular gas provides a homogeneous thermal bath with appropriate viscosity $\eta_g$, such that the Reynolds numbers remain very small ($Re \sim 10^{-4}$) and the Stokes number, $T_g^*$, as well as $\gamma/\nu$, are all close to unity. This ensures that the molecular gas effectively thermalises the granular particles, while the effect of granular collisions remains significant.

Once the particles composing the system are selected, an experimental setup to investigate segregation of intruders in a granular gas immersed in an interstitial fluid can be designed based on previous studies \cite[]{MLNJ01,NSK03,SSK04,WZXS08,CPSK10,PGM14}. The setup can consist of a transparent container with a porous base to allow fluid flow while maintaining particle collisions, filled with small granular particles and a single intruder whose mass and size can be adjusted. The interstitial fluid, either a gas or a liquid, occupies the voids between particles, enabling the study of drag and thermalisation effects. The container is subjected to controlled vertical or horizontal vibrations with adjustable amplitude, and the position of the intruder is tracked using high-speed imaging. By varying the intruder’s mass, size, and the properties of the fluid, this setup allows for systematic measurement of segregation phenomena, including the influence of mass ratio and interstitial fluid on the BNE/RBNE transition.

}

Finally, it is important to remark that to facilitate the solution of the integral equations verifying the mixture, we have considered the tracer limit where one of the species is present at a negligible concentration. A possible extension of this work is to generalise diffusion to a binary mixture with arbitrary concentrations. In addition, it may be interesting to study the role of density in the diffusion and segregation diagrams using the Enskog kinetic equation which applies to moderate densities. \vicente{Moreover, revisiting the problem within a simplified random-walk framework could shed light on the underlying mechanisms responsible for the observed non-monotonicities and the crossing of curves in the tracer diffusion coefficient}. These studies will be developed in future projects.

\vspace{0.5cm}

 \noindent \textbf{Acknowledgements.} 
 
 VG acknowledges financial support from Grant No. PID2024-156352NB-I00 and from Grant No. GR24022 funded by Junta de Extremadura (Spain) and by European Regional Development Fund (ERDF) ``A way of making Europe.''
 

\noindent\textbf{Declaration of interests.} The authors report no conflict of interest.\\

\noindent\textbf{Author ORCIDs.}
\\ Rub\'en G\'omez Gonz\'alez: https://orcid.org/0000-0002-5906-5031
\\Vicente Garz\'o: https://orcid.org/0000-0001-6531-9328

\appendix
\section{Determination of the kurtosis $c_0$}
\label{appA}

In this appendix we give some details on the determination of the kurtosis $c_0$. It is defined by Eq.\ (4.13). To obtain $c_0$ one needs first to estimate the (reduced) cooling rates $\zeta_0^*$ and $ \zeta_{0g}^*$ defined by Eq.\ (4.10). As usual, to get these cooling rates one replaces $f$  by its first Sonine approximation
\beq
\label{a1}
f_0(\mathbf{v})\to f_{0,\text{M}}(\mathbf{v})\Bigg\{1+\frac{c_0}{2}\Bigg[\Big(\frac{m_0v^2}{2T_0}\Big)^2-(d+2)\frac{m_0 v^2}{2T_0}+\frac{d(d+2)}{4}\Bigg]\Bigg\},
\eeq
where the Maxwellian distribution $f_{0,\text{M}}$ is defined in Eq.\ (4.12). Neglecting nonlinear terms in $c_0$, $\zeta_0^*$ and $ \zeta_{0g}^*$ can be written as \cite[]{GGG21}
\beq
\label{a2}
\zeta_0^*=\zeta_{00}+\zeta_{01}c_0+\zeta_{02}c, \quad \zeta_{0g}^*=\zeta_{0g,0}+\zeta_{0g,1}c_0,
\eeq
where the expression of the kurtosis $c$ of the granular gas is given by Eq.\ (3.19) of \cite{GG22} while the forms of $\zeta_{00}$ and $\zeta_{0g,0}$ are given by Eqs.\ (4.15) and (4.16), respectively. The expressions of $\zeta_{01}$, $\zeta_{02}$, and $\zeta_{0g,1}$ are given by
\beq
\label{a3}
\zeta_{01}=\frac{\pi^{(d-1)/2}}{2d\Gamma\left(\frac{d}{2}\right)}
\left(\frac{\sigma'}{\sigma}\right)^{d-1}\mu'
\frac{(1+\theta_0)^{-3/2}}{\theta_0^{1/2}}(1+\alpha_{0}) \Big[3+4\theta_0-\frac{3}{2}\mu'(1+\alpha_{0})(1+\theta_0) \Big],
\eeq
\beq
\label{a4}
\zeta_{02}=-\frac{\pi^{(d-1)/2}}{2d\Gamma\left(\frac{d}{2}\right)}
\left(\frac{\sigma'}{\sigma}\right)^{d-1}\mu'
\left(\frac{1+\theta_0}{\theta_0}\right)^{-3/2}(1+\alpha_{0})
\Big[1+\frac{3}{2}\mu'(1+\alpha_{0})(1+\theta_0) \Big],
\eeq
\beq
\label{a5}
\zeta_{0g,1}=\frac{1}{8}\mu_{g0}x_0^{-3}\Big[x_0^2\left(4-3\mu_{g0}\right)-\mu_{g0}\Big]
\left(\mu_{0g}\frac{T_0}{T_g}\right)^{1/2}\gamma_0^*.
\eeq

To determine $c_0$ one has to obtain also the fourth-degree collisional moments
\beq
\label{a6}
\Lambda_0=\int d\mathbf{v}\; v^4\; J_0[f_0,f], \quad \Lambda_{0g}=\int d\mathbf{v}\; v^4\; J_{0g}[f_0,f_g].
\eeq
These moments have been computed in previous works \cite[]{GGG21} by neglecting nonlinear terms in $c$ and $c_0$.
Their dimensionless forms
\beq
\label{a7}
\Lambda_0^*=\frac{m_0^2}{n_0 T_0^2 \nu}\Lambda_0, \quad \Lambda_{0g}^*=\frac{m_0^2}{n_0 T_0^2 \nu}\Lambda_{0g},
\eeq
can be written as
\beq
\label{a8}
\Lambda_0^*=\Lambda_{00}+\Lambda_{01}c_0+\Lambda_{02}c, \quad \Lambda_{0g}^*=\Lambda_{0g,0}+\Lambda_{0g,1}c_0.
\eeq
Here,
\begin{eqnarray}
\label{a9}
\Lambda_{00} &=&\frac{4\pi^{(d-1)/2}}{\Gamma\left(\frac{d}{2}\right)} \left(\frac{\sigma'}{\sigma}\right)^{d-1}\mu'
\left[\theta_0(1+\theta_0)\right]^{-1/2}(1+\al_0)\Big\{-2\left[d+3+(d+2)\theta_0\right] \nonumber\\
& &
+\mu'\left(1+\alpha_{0}\right) \left( 1+\theta_0\right)
\left(11+ d+\frac{d^2+5d+6}{d+3} \theta_0 \right)-8\mu^{'2}\left(1+\alpha_{0}\right)^{2}\left(1+\theta_0 \right)^{2} \nonumber\\
& & +2\mu^{'3}\left(
1+\alpha_{0}\right)^{3}\left( 1+\theta_0 \right)^{3}\Big\},
\end{eqnarray}
\begin{eqnarray}
\label{a10}
\Lambda_{01} &=&
\frac{\pi^{(d-1)/2}}{2\Gamma\left(\frac{d}{2}\right)}\left(\frac{\sigma'}{\sigma}\right)^{d-1}\mu' \theta_0^{-1/2}\left(1+\theta_0\right)^{-5/2}
\left(1+\alpha_{0}\right)\Big\{-2\big[ 45+15d+(114+39d)\theta_0\nonumber\\
& & +(88+32d)\theta_0^{2}+(16+8d)\theta_0^{3}\big]+3\mu'\left(1+\alpha_{0}\right) \left(1+\theta_0 \right) \left[55+5d+9(10+d)\theta_0\right.\nonumber\\
& & \left.+4(8+d)\theta_0
^{2}\right]-24\mu^{'2}\left(1+\alpha_{0}\right)^{2}\left(1+\theta_0 \right)^{2}\left(5+4\theta_0
\right)+30\mu^{'3}\left(1+\alpha_{0}\right)^{3}\left(1+\theta_0\right)^{3}\Big\},
\nonumber\\
\end{eqnarray}
\beqa
\label{a11}
\Lambda_{02}&=&\frac{\pi^{(d-1)/2}}{2\Gamma\left(\frac{d}{2}\right)}\left(\frac{\sigma'}{\sigma}\right)^{d-1}\mu'
\theta_0^{3/2}\left(1+\theta_0\right)^{-5/2} \left(
1+\alpha_{0}\right) \Big\{ 2\left[d-1+(d+2)\theta_0\right]\nonumber\\
& & +3\mu'\left(1+\alpha_{0}\right) \left(1+\theta_0\right)
\left[d-1+(d+2)\theta_0\right]
-24\mu^{'2}\left( 1+\alpha_{0}\right)^{2}\left(1+\theta_0 \right)^{2}\nonumber\\
& &
+30\mu^{'3}\left(1+\alpha_{0}\right)^{3}\left(1+\theta_0\right)^{3}\Big\},
\eeqa
\beq
\label{a12}
\Lambda_{0g,0}=4d x_0^{-1}\left(x_0^2-1\right)\left[8 \mu_{g0} x_0^4+x_0^2\left(d+2-8\mu_{g0}\right)+\mu_{g0}\right]\Big(\frac{\mu_{0g} T_0}{T_g}\Big)^{1/2}\gamma_0^*,
\eeq
\beqa
\label{a13}
\Lambda_{0g,1}&=&\frac{d}{2} x_0^{-5}\Bigg\{4x_0^6\Big[30\mu_{g0}^3-48\mu_{g0}^2+3(d+8)\mu_{g0}-2(d+2)\Big]+\mu_{g0} x_0^4 \Big[-48\mu_{g0}^2\nonumber\\
& &+3(d+26)\mu_{g0}-8(d+5)\Big] +\mu_{g0}^2 x_0^2\left(d+14-9\mu_{g0}\right)-3\mu_{g0}^3\Bigg\}\Big(\frac{\mu_{0g} T_0}{T_g}\Big)^{1/2}\gamma_0^*.
\nonumber\\
\eeqa

In the steady state, the temperature ratio $\chi_0=T_0/T$ and the cumulant $c_0$ are obtained from the constraints
\beq
\label{a14}
\zeta_{0}^*+\zeta_{0g}^*=0, \quad \Lambda_{0}^*+\Lambda_{0g}^*=0.
\eeq
Inserting \eqref{a2} and \eqref{a8} into \eqref{a14} yields the set of coupled equations:
\beq
\label{a15}
\zeta_{00}^*+\zeta_{02}^*c+\zeta_{0g,0}^*+\left(\zeta_{01}^*+\zeta_{0g,1}^*\right)c_0=0,
\eeq
\beq
\label{a16}
\Lambda_{00}^*+\Lambda_{02}^*c+\Lambda_{0g,0}^*+\left(\Lambda_{01}^*+\Lambda_{0g,1}^*\right)c_0=0.
\eeq
The numerical solution to this set provides the dependence of $c_0$ on the parameter space of the system.

Another equivalent way of obtaining $\chi_0$ and $c_0$ is to eliminate first $c_0$ in Eqs.\ \eqref{a15} and \eqref{a16} to achieve the equation
\beq
\label{a17}
\left(\zeta_{00}+\zeta_{02}c+\zeta_{0g,0}\right)\left(\Lambda_{01}+\Lambda_{0g,1}\right)=\left(\zeta_{01}+\zeta_{0g,1}\right)
\left(\Lambda_{00}+\Lambda_{02}c+\Lambda_{0g,0}\right).
\eeq
The solution to Eq.\ \eqref{a17} gives $\chi_0$ for given values of the parameters of the system. Once $\chi_0$ is known, the expression of $c_0$ is
\beq
\label{a18}
c_0=-\frac{\zeta_{00}+\zeta_{02}c+\zeta_{0g,0}}{\zeta_{01}+\zeta_{0g,1}}=-\frac{\Lambda_{00}+\Lambda_{02}c+\Lambda_{0g,0}}
{\Lambda_{01}+\Lambda_{0g,1}}.
\eeq

\section{First-order distribution function for intruders}
\label{appB}

To first order in gradients, the distribution $f_0^{(1)}$ verifies the kinetic equation
\beqa
\label{b1}
\partial_t^{(0)}f_0^{(1)}-J_0[f_0^{(1)},f^{(0)}]-J_{0g}[f_0^{(1)},f_g^{(0)}]&=&-\left(D_t^{(1)}+\mathbf{V}\cdot \nabla+\mathbf{g}\cdot  \frac{\partial}{\partial \mathbf{v}}\right)f_0^{(0)}
\nonumber\\
& & +J_0[f_0^{(0)},f^{(1)}]-\frac{m_g}{T_g}\Delta \mathbf{U}\cdot J_{0g}[f_0^{(0)},\mathbf{V}f_g^{(0)}],
\nonumber\\
\eeqa
where $D_t^{(1)}\equiv \partial_t^{(1)}+\mathbf{V}\cdot \nabla$. Moreover, the first-order distribution $f^{(1)}$ of the granular gas is given by \cite[]{GG22}
\beqa
\label{b2}
f^{(1)}(\mathbf{V})&=&\boldsymbol{\mathcal{A}}(\mathbf{V})\cdot \nabla T+\boldsymbol{\mathcal{B}}(\mathbf{V})\cdot \nabla n+\mathcal{C}_{ij}(\mathbf{V})\left(\partial_iU_j+\partial_j U_i-\frac{2}{d}\delta_{ij}\nabla\cdot \mathbf{U}\right)\nonumber\\
& & +\mathcal{D}(\mathbf{V}) \nabla \cdot \mathbf{U}+\boldsymbol{\varepsilon}(\mathbf{V})\cdot \Delta \mathbf{U},
\eeqa
where the unknowns $(\boldsymbol{\mathcal{A}}, \boldsymbol{\mathcal{B}},\mathcal{C}_{ij},\mathcal{D},\boldsymbol{\varepsilon})$ have been estimated by \cite{GG22} in their corresponding leading Sonine approximations. The balance equations to first order are
\beq
\label{b3}
D_t^{(1)}n=-n\nabla \cdot \mathbf{U}, \quad D_t^{(1)}n_0=-n_0\nabla \cdot \mathbf{U}, \quad D_t^{(1)}T=-\frac{2T}{d}\nabla\cdot \mathbf{U}-T\left(\zeta^{(1)}+\zeta_g^{(1)}\right),
\eeq
\beq
\label{b4}
D_t^{(1)}\mathbf{U}=-\rho^{-1}\nabla p+\mathbf{g}-\rho^{-1}\xi \Delta \mathbf{U}+\rho^{-1}\boldsymbol{\mathcal{K}}[f^{(1)}],
\eeq
where
\beq
\label{b5}
\boldsymbol{\mathcal{K}}[X]=\int d\mathbf{v}\; m \mathbf{V}\; J_g[X,\mathbf{V}f_g^{(0)}],
\eeq
\beq
\label{b6}
\xi=\frac{1}{d}\frac{m_g}{T_g}\int d\mathbf{v}m\mathbf{V}\cdot J_g[f^{(0)},\mathbf{V}f_g^{(0)}].
\eeq
Note that the scalars $\zeta^{(1)}$ and $\zeta_g^{(1)}$ can be only proportional to the divergence of the flow velocity field $\nabla \cdot \mathbf{U}$. Therefore,
\beq
\label{b7}
\zeta^{(1)}=\zeta_U \nabla \cdot \mathbf{U}, \quad \zeta_g^{(1)}=\zeta_{Ug} \nabla \cdot \mathbf{U}.
\eeq

Equation \eqref{b1} can be more explicitly written when one takes into account Eqs.\ \eqref{b3}--\eqref{b7} as
\beqa
\label{b8}
& & \partial_t^{(0)}f_0^{(1)}-J_0[f_0^{(1)},f^{(0)}]-J_{0g}[f_0^{(1)},f_g^{(0)}]=\mathbf{A}_0\cdot \nabla T+\mathbf{B}_0\cdot \nabla n+\mathbf{C}_0\cdot \nabla n_0+D'_0 \nabla \cdot \mathbf{U}\nonumber\\
& & +D_{0,ij}\left(\partial_iU_j+\partial_j U_i-\frac{2}{d}\delta_{ij}\nabla\cdot \mathbf{U}\right)+\mathbf{E}_0 \cdot \Delta \mathbf{U},
\eeqa
where
\beq
\label{b9}
\mathbf{A}_0(\mathbf{V})=-\mathbf{V}\frac{\partial f_0^{(0)}}{\partial T}-\frac{p}{\rho T}\frac{\partial f_0^{(0)}}{\partial \mathbf{V}}+J_0[f_0^{(0)},\boldsymbol{\mathcal A}], \quad
\mathbf{B}_0(\mathbf{V})=-\mathbf{V} \frac{\partial f_0^{(0)}}{\partial n}-\frac{T}{\rho}\frac{\partial f_0^{(0)}}{\partial \mathbf{V}}+J_0[f_0^{(0)},\boldsymbol{\mathcal B}],
\eeq
\beq
\label{b10}
\mathbf{C}_0(\mathbf{V})=-\mathbf{V} \frac{\partial f_0^{(0)}}{\partial n_0}
,\quad D_{0,ij}(\mathbf{V})=V_i\frac{\partial f_0^{(0)}}{\partial V_j}+J_0[f_0^{(0)},\mathcal{C}_{ij}],
\eeq
\beq
\label{b11}
D_0'(\mathbf{V})=\frac{1}{d}\frac{\partial}{\partial \mathbf{V}}\cdot \left(\mathbf{V}f_0^{(0)}\right)+\frac{2}{d}
T\frac{\partial f_0^{(0)}}{\partial T}-f_0^{(0)}+n_0\frac{\partial f_0^{(0)}}{\partial n_0}+T\left(\zeta_U+\zeta_{Ug}\right)
\frac{\partial f_0^{(0)}}{\partial T}+J_0[f_0^{(0)},\mathcal {D}],
\eeq
\beq
\label{b12}
\mathbf{E}_0(\mathbf{V})=-\rho^{-1} \frac{\partial f_0^{(0)}}{\partial \mathbf{V}}\xi-\frac{m_g}{T_g} J_{0g}[f_0^{(0)},\mathbf{V}f_g^{(0)}]+J_0[f_0^{(0)},\boldsymbol{\mathcal \varepsilon}].
\eeq
Note that the external field does not appear in the kinetic equation \eqref{b8}. This is due to the particular form of the gravitational force.

The solution to Eq.\ \eqref{b8} is
\beqa
\label{b13}
f_0^{(1)}(\mathbf{V})&=&\boldsymbol{\mathcal{A}}_0(\mathbf{V})\cdot \nabla T+\boldsymbol{\mathcal{B}}_0(\mathbf{V})\cdot \nabla n+\boldsymbol{\mathcal{C}}_0(\mathbf{V})\cdot \nabla n_0+\mathcal{D}_0'(\mathbf{V}) \nabla \cdot \mathbf{U}\nonumber\\
& &+\mathcal{D}_{0,ij}(\mathbf{V})\left(\partial_iU_j+\partial_j U_i-\frac{2}{d}\delta_{ij}\nabla\cdot \mathbf{U}\right)
+\boldsymbol{\varepsilon}_0(\mathbf{V})\cdot \Delta \mathbf{U}.
\eeqa
Since the mass flux $\mathbf{j}_0^{(1)}$ is a vector, it can be only coupled to the gradients $\nabla T$, $\nabla n$, and $\nabla n_0$ and the term $\Delta \mathbf{U}$. Its constitutive equation is
\beq
\label{b14}
\mathbf{j}_0^{(1)}=-\frac{m_0^2}{\rho}D_0 \nabla n_0-\frac{m m_0}{\rho}D \nabla n-\frac{\rho}{T}D_T \nabla T-D_0^U \Delta \mathbf{U},
\eeq
where the diffusion transport coefficients are defined as
\beq
\label{b15}
D_T=-\frac{m_0 T}{d\rho}\int d\mathbf{v}\; \mathbf{V}\cdot \boldsymbol{\mathcal{A}}_0(\mathbf{V}),
\eeq
\beq
\label{b16}
D=-\frac{n}{d}\int d\mathbf{v}\; \mathbf{V}\cdot \boldsymbol{\mathcal{B}}_0(\mathbf{V}),
\eeq
\beq
\label{b17}
D_0=-\frac{\rho}{dm_0}\int d\mathbf{v}\; \mathbf{V}\cdot \boldsymbol{\mathcal{C}}_0(\mathbf{V}),
\eeq
\beq
\label{b18}
D_0^U=-\frac{m_0}{d}\int d\mathbf{v}\; \mathbf{V}\cdot \boldsymbol{\mathcal{\varepsilon}}_0(\mathbf{V}).
\eeq

Substitution of the expression \eqref{b13} into Eq.\ \eqref{b8} allows us to obtain the set of coupled linear integral equations obeying the unknowns $(\boldsymbol{\mathcal{A}}_0, \boldsymbol{\mathcal{B}}_0,\boldsymbol{\mathcal{C}}_0,\mathcal{D}_{0,ij},\mathcal{D}_0',\boldsymbol{\varepsilon}_0)$. In the case of the quantities involved in the determination of the mass flux, the integral equations are given by
\beqa
\label{b19}
& & -\left(\zeta^{(0)}+\zeta_g^{(0)}\right)T\partial_T \mathcal{A}_{0,i}-\left[\frac{3}{2}\left(\zeta^{(0)}+\zeta_g^{(0)}\right)+\zeta_g^{(0)} \chi \frac{\partial}{\partial \chi}\ln \zeta_g^*\right]\mathcal{A}_{0,i}-J_0[\mathcal{A}_{0,i},f^{(0)}]\nonumber\\
& & -J_{0g}[\mathcal{A}_{0,i},f_g^{(0)}]
=A_{0,i}+\rho^{-1}\frac{\partial f_0^{(0)}}{\partial V_j}\mathcal{K}_j[\mathcal{A}_{i}],
\eeqa
\beqa
\label{b20}
& & -\left(\zeta^{(0)}+\zeta_g^{(0)}\right)T\partial_T \mathcal{B}_{0,i}-J_0[\mathcal{B}_{0,i},f^{(0)}]-J_{0g}[\mathcal{B}_{0,i},f_g^{(0)}]
=B_{0,i}+\rho^{-1}\frac{\partial f_0^{(0)}}{\partial V_j}\mathcal{K}_j[\mathcal{B}_{i}]\nonumber\\
& & +\frac{T}{n}\left[\zeta^{(0)}+\zeta_g^{(0)}\left(1-\epsilon \frac{\partial}{\partial \epsilon}\ln \zeta_g^*\right)\right]\mathcal{A}_{0,i},
\eeqa
\beq
\label{b21}
-\left(\zeta^{(0)}+\zeta_g^{(0)}\right)T\partial_T \mathcal{C}_{0,i}-J_0[\mathcal{C}_{0,i},f^{(0)}]-J_{0g}[\mathcal{C}_{0,i},f_g^{(0)}]
=C_{0,i},
\eeq
\beq
\label{b22}
-\left(\zeta^{(0)}+\zeta_g^{(0)}\right)T\partial_T \mathcal{\varepsilon}_{0,i}-J_0[\mathcal{\varepsilon}_{0,i},f^{(0)}]-J_{0g}[\mathcal{\varepsilon}_{0,i},f_g^{(0)}]
=E_{0,i}+\rho^{-1}\frac{\partial f_0^{(0)}}{\partial V_j}\mathcal{K}_j[\mathcal{\varepsilon}_{i}].
\eeq
In Eqs.\ \eqref{b19}--\eqref{b22}, $\zeta^{(0)}$ and $\zeta_g^{(0)}$ are defined by Eqs.\ (4.5) when one makes the replacements $f\to f^{(0)}$ and $f_g \to f_g^{(0)}$, respectively, and approximates $f^{(0)}$ and $f_0^{(0)}$ by their Maxwellian distribution forms (4.4) and (4.12), respectively. In addition, upon obtaining Eqs.\ \eqref{b19}--\eqref{b22}, use has been made of the result
\beqa
\label{b23}
\partial_t^{(0)}\nabla T&=&\nabla \partial_t^{(0)} T=-\nabla \left[T \left(\zeta^{(0)}+\zeta_g^{(0)}\right)\right]
=-\frac{T}{n}\left[\zeta^{(0)}+\zeta_g^{(0)}\left(1-\epsilon \frac{\partial}{\partial \epsilon}\ln \zeta_g^*\right)\right]\nabla n\nonumber\\
& & -
\left[\frac{3}{2}\left(\zeta^{(0)}+\zeta_g^{(0)}\right)+\zeta_g^{(0)} \chi \frac{\partial}{\partial \chi}\ln \zeta_g^*\right]\nabla T.
\eeqa

The linear integral equations obeying the diffusion transport coefficients can be derived from Eqs.\ \eqref{b19}--\eqref{b22} when one takes into account their definitions \eqref{b15}--\eqref{b18}. However,
as occurs in the case of monocomponent granular suspensions (granular particles immersed in a molecular gas) \cite[]{GG22}, the solution to these integral equations for general \emph{unsteady} conditions requires one to numerically solve them. Thus, to achieve analytical expressions for the diffusion transport coefficients we assume steady-state  conditions. This means that the constraint $\zeta^{(0)}+\zeta_g^{(0)}=0$ applies locally and hence, the first term on the left-hand side of Eqs.\ \eqref{b19}--\eqref{b22} vanishes. This yields the set of coupled integral equations
\beq
\label{b24}
\beta \gamma \mathcal{A}_{0,i}-J_0[\mathcal{A}_{0,i},f^{(0)}]-J_{0g}[\mathcal{A}_{0,i},f_g^{(0)}]
=A_{0,i}+\rho^{-1}\frac{\partial f_0^{(0)}}{\partial V_j}\mathcal{K}_j[\mathcal{A}_{i}],
\eeq
\beq
\label{b25}
-J_0[\mathcal{B}_{0,i},f^{(0)}]-J_{0g}[\mathcal{B}_{0,i},f_g^{(0)}]
=B_{0,i}+\rho^{-1}\frac{\partial f_0^{(0)}}{\partial V_j}\mathcal{K}_j[\mathcal{B}_{i}]-\frac{T}{n}\zeta_g\mathcal{A}_{0,i},
\eeq
\beq
\label{b26}
-J_0[\mathcal{C}_{0,i},f^{(0)}]-J_{0g}[\mathcal{C}_{0,i},f_g^{(0)}]
=C_{0,i},
\eeq
\beq
\label{b27}
-J_0[\mathcal{\varepsilon}_{0,i},f^{(0)}]-J_{0g}[\mathcal{\varepsilon}_{0,i},f_g^{(0)}]
=E_{0,i}+\rho^{-1}\frac{\partial f_0^{(0)}}{\partial V_j}\mathcal{K}_j[\mathcal{\varepsilon}_{i}].
\eeq
Upon obtaining Eqs.\ \eqref{b24} and \eqref{b25}, use has been made of the results
\beq
\label{b28}
-\zeta_g^{(0)} \chi \frac{\partial \ln \zeta_g^*}{\partial \chi}=\beta \gamma, \quad \beta=\left(x^{-1}-3x\right)\mu^{3/2}\chi^{-1/2},
\eeq
\beq
\label{b29}
\zeta_g^{(0)} \epsilon \frac{\partial \ln \zeta_g^*}{\partial \epsilon}=\zeta_g^{(0)}=-\zeta^{(0)},
\eeq
where $\epsilon$ is defined by Eq.\ (4.7).

\section{Leading Sonine approximations to the diffusion transport coefficients}
\label{appc}

The integral equations \eqref{b24}--\eqref{b27} are still exact. However, to get the diffusion transport coefficients one has to solve the above integral equations as well as to know the zeroth-order distributions $f^{(0)}$ and $f_0^{(0)}$. On the other hand, as show in the results obtained for the HSS, the non-Gaussian corrections to the above distributions (measured by the kurtosis $c$ and $c_0$) are in general very small. Thus, we approximate $f^{(0)}$ and $f_0^{(0)}$ by the Maxwellian distributions (4.4) and (4.12), respectively. With respect to the functions $\boldsymbol{\mathcal{A}}_0$, $\boldsymbol{\mathcal{B}}_0$, $\boldsymbol{\mathcal{C}}_0$, and $\boldsymbol{\varepsilon}_0$, as usual we consider the leading term in a series expansion of these quantities in Sonine polynomials. At this level of approximation, the quantities associated with the granular gas vanish (i.e., $\boldsymbol{\mathcal{A}}\to 0$ and $\boldsymbol{\mathcal{B}}\to 0$) while the quantities of the tracer species are approximated by
\beq
\label{c1}
\boldsymbol{\mathcal{A}}_0(\mathbf{V})\to -f_{0,\text{M}}\mathbf{V}\frac{\rho}{Tn_0T_0}D_T, \quad
\boldsymbol{\mathcal{B}}_0(\mathbf{V})\to -f_{0,\text{M}}\mathbf{V}\frac{m_0}{n n_0T_0}D,
\eeq
\beq
\label{c2}
\boldsymbol{\mathcal{C}}_0(\mathbf{V})\to -f_{0,\text{M}}\mathbf{V}\frac{m_0^2}{\rho n_0T_0}D_0, \quad
\boldsymbol{\mathcal{\varepsilon}}_0(\mathbf{V})\to -f_{0,\text{M}}\mathbf{V}\frac{D_0^U}{n_0T_0}.
\eeq
To determine $D_0$, $D$, $D_T$, and $D_0^U$, we substitute first $\boldsymbol{\mathcal{A}}_0$, $\boldsymbol{\mathcal{B}}_0$, $\boldsymbol{\mathcal{C}}_0$, and $\boldsymbol{\varepsilon}_0$ by their leading Sonine approximations \eqref{c1} and \eqref{c2} in Eqs.\ \eqref{b24}--\eqref{b27}, respectively. Then we multiply these equations by $m_0\mathbf{V}$ and integrate over velocity. Let us evaluate each transport coefficient separately.

\subsection{Thermal diffusion coefficient $D_T$}

As said before, multiplying both sides of Eq.\ \eqref{b24} and integrating over $\mathbf{v}$ one gets
\beq
\label{c3}
\left(\beta \gamma^*+\nu_D^*+\widetilde{\nu}_D\gamma_0^*\right)D_T^*=\tau_0-\frac{m_0}{m}+T\frac{\partial \tau_0}{\partial T},
\eeq
where $D_T^*=(\rho \nu/n_0 T)D_T$, $\tau_0=T_0/T$,
\beq
\label{c4}
\nu_D^*=\frac{2\pi^{(d-1)/2}}{d\Gamma\left(\frac{d}{2}\right)}\left(\frac{\sigma'}{\sigma}\right)^{d-1}\mu' \left(\frac{1+\theta_0}{\theta_0}\right)^{1/2}(1+\al_0),
\eeq
\beq
\label{c5}
\widetilde{\nu}_D=\left(\frac{m_0T_0}{m_g T_g}\right)^{1/2}\mu_{g0}\left(1+\theta_{0g}\right)^{1/2}.
\eeq
In Eqs.\ \eqref{c3}--\eqref{c5}, we have introduced the quantities
\beq
\label{c6}
\theta_0=\frac{m_0T}{m T_0}, \quad \theta_{0g}=\frac{m_0T_g}{m_g T_0},
\eeq
and use has been made of the results \cite[]{GM07}
\beq
\label{c7}
\int d\mathbf{v}\; m_0 \mathbf{V}\cdot \mathbf{A}_0=dn_0\left(\frac{m_0}{m}-\tau_0\right)-d n_0 T\frac{\partial \tau_0}{\partial T},
\eeq
\beq
\label{c8}
\int d\mathbf{v}\; m_0 \mathbf{V}\cdot J_0[\mathbf{V}f_{0,\text{M}}, f^{(0)}]=-\frac{2\pi^{(d-1)/2}}{\Gamma\left(\frac{d}{2}\right)}\left(\frac{\sigma'}{\sigma}\right)^{d-1}\mu'\nu n_0 T_0 \left(\frac{1+\theta_0}{\theta_0}\right)^{1/2}(1+\al_0),
\eeq
\beq
\label{c9}
\int d\mathbf{v}\; m_0 \mathbf{V}\cdot J_{0g}[\mathbf{V}f_{0,\text{M}}, f_g^{(0)}]=-d n_0 T_0 \left(\frac{m_0T_0}{m_g T_g}\right)^{1/2}\mu_{g0}\left(1+\theta_{0g}\right)^{1/2}\gamma_0.
\eeq
The solution to Eq.\ \eqref{c3} is
\beq
\label{c10}
D_T^*=\frac{\tau_0-\frac{m_0}{m}+\chi\frac{\partial \tau_0}{\partial \chi}}{\beta \gamma^*+\nu_D^*+\widetilde{\nu}_D\gamma_0^*},
\eeq
where we have taken into account the identity $T\partial_T \tau_0=\chi\partial_\chi \tau_0$.

\subsection{Mutual diffusion coefficient $D$}

Proceeding in a similar way as in the case of $D_T$, Eq.\eqref{b25} yields the result
\beq
\label{c11}
\left(\nu_D^*+\widetilde{\nu}_D\gamma_0^*\right)D^*=\zeta^* D_T^*-\frac{m_0}{m}+\phi\frac{\partial \tau_0}{\partial \phi},
\eeq
where $D^*=(m_0\nu/n_0 T)D$ and use has been made of the result
\beq
\label{c12}
\int d\mathbf{v}\; m_0 \mathbf{V}\cdot \mathbf{B}_0=-dx_0 n T\frac{\partial \tau_0}{\partial n}+d x_0 \frac{m_0}{m}T.
\eeq
The solution to Eq.\ \eqref{c12} is simply given by
\beq
\label{c13}
D^*=\frac{\zeta^* D_T^*-\frac{m_0}{m}+\phi\frac{\partial \tau_0}{\partial \phi}}{\nu_D^*+\widetilde{\nu}_D\gamma_0^*}.
\eeq

\subsection{Tracer diffusion coefficient $D_0$}

The (reduced) tracer diffusion coefficient $D_0^*=[m_0^2 \nu/(m n T)]D_0$ can be determined from Eq.\ \eqref{b26} as
\beq
\label{c14}
D_0^*=\frac{\tau_0}{\nu_D^*+\widetilde{\nu}_D\gamma_0^*},
\eeq
where use has been made of the result
\beq
\label{c15}
\int d\mathbf{v}\; m_0 \mathbf{V}\cdot \mathbf{C}_0=-d T_0.
\eeq

\subsection{Velocity diffusion coefficient $D_0^U$}

In dimensionless form, the (reduced) diffusion coefficient $D_0^{*U}=D_0^{U}/(m_0 n_0)$ can be obtained from Eq.\ \eqref{b27} as
\beq
\label{c16}
D_0^{U*}=\frac{\xi_0^*-\xi^*}{\nu_D^*+\widetilde{\nu}_D\gamma_0^*},
\eeq
where
\beq
\label{c17}
\xi_0^*=\frac{\xi_0}{\rho_0\nu}=\mu_{0g}\theta_{0g}^{-1/2}\left(1+\theta_{0g}\right)^{1/2}\gamma_0^*,
\eeq
\beq
\label{c18}
\xi^*=\frac{\xi}{\rho \nu}=\mu\theta^{-1/2}\left(1+\theta\right)^{1/2}\gamma^*,
\eeq
where $\rho_0=m_0n_0$, $\rho=m n$, and $\theta=m T_g/(m_g T)$. Upon obtaining Eq.\ \eqref{c16}, we have taken into account the results
\beq
\label{c19}
\xi_0=\frac{1}{d}\frac{m_g}{T_g}\int d\mathbf{v}\; m_0 \mathbf{V}\cdot J_{0g}[f_0^{(0)},\mathbf{V}f_g^{(0)}]=\rho_0\mu_{0g}\theta_{0g}^{-1/2}\left(1+\theta_{0g}\right)^{1/2}\gamma_0,
\eeq
\beq
\label{c20}
\xi=\frac{1}{d}\frac{m_g}{T_g}\int d\mathbf{v}\; m_0 \mathbf{V}\cdot J[f^{(0)},\mathbf{V}f_g^{(0)}]=\rho \mu\theta^{-1/2}\left(1+\theta\right)^{1/2}\gamma,
\eeq
\beq
\label{c21}
\int d\mathbf{v}\; m_0 \mathbf{V}\cdot \mathbf{E}_0=d\left(\frac{\rho_0}{\rho}\xi-\xi_0\right).
\eeq

\section{Expressions of $\kappa^*$ and $\overline{\mu}^*$}
\label{appD}

The expression of $\kappa^*$ is given by \cite[]{GG22}
\beq
\label{d1}
\kappa^*=\frac{d-1}{d}\frac{1}{\nu_\kappa^*+K\left(\widetilde{\nu}_\kappa+\beta\right)\gamma^*},
\eeq
where $\beta$ is defined by Eq.\ (6.4), 
\beq
\label{d2}
K=\sqrt{2} \frac{(d+2)\Gamma\Big(\frac{d}{2}\Big)}{8\pi^{(d-1)/2}},
\eeq
\beq
\label{d3}
\nu_\kappa^*=\frac{1+\al}{d}\left[\frac{d-1}{2}+\frac{3}{16}(d+8)(1-\al)\right](1+\al),
\eeq
and 
\beq
\label{d4}
\widetilde{\nu}_\kappa=\frac{1}{2(d+2)}\mu \frac{\theta}{1+\theta} \Big[G-(d+2)\frac{1+\theta}{\theta}F\Big].
\eeq
In Eq.\ \eqref{d4}, we have introduced the quantities
\beqa
\label{d5}
F&=&(d+2)(2 \delta +1)+4(d-1)\mu_g \delta \theta^{-1}(1+\theta)+3(d+3)\delta^2\theta^{-1}+(d+3)\mu_g^2 \theta^{-1}(1+\theta)^2\nonumber\\
& & -(d+2)\theta^{-1}(1+\theta),
\eeqa
\beqa
\label{d6}
G&=&(d+3)\mu_g^2 \theta^{-2}(1+\theta)^2\left[d+5+(d+2)\theta\right]-\mu_g(1+\theta)\Big\{4(1-d)\delta \theta^{-2}\left[d+5+(d+2)\theta\right]
\nonumber\\
& & -8(d-1)\theta^{-1}\Big\}+3(d+3)\delta^2\theta^{-2}\left[d+5+(d+2)\theta\right] +2\delta\theta^{-1}\left[24+11d+d^2\right.\nonumber\\
& & \left.+(d+2)^2\theta\right]+(d+2)\theta^{-1} \left[d+3+(d+8)\theta\right]-(d+2)\theta^{-2}(1+\theta)\left[d+3+(d+2)\theta\right],
\nonumber\\
\eeqa
where $\delta\equiv \mu-\mu_g \theta$.

The expression of $\overline{\mu}^*$ is \cite[]{GG22}
\beq
\label{d7}
\overline{\mu}^*=\frac{K\kappa^* \zeta^*}{\nu_\kappa^*+K\widetilde{\nu}_\kappa \gamma^*}.
\eeq

\section{Derivatives of the temperature ratio $\tau_0$ with respect to $\chi$ and $\phi$}
\label{appE}

To determine the (reduced) thermal diffusion $D^{T*}$ and mutual diffusion $D^*$ coefficients one needs to evaluate the derivatives $\tau_{\chi,0}\equiv \partial \tau_0/\partial \chi$ and $\tau_{\phi,0}\equiv \partial \tau_0/\partial \phi$ in the steady state. To determine these derivatives, we start from the relations $\partial_t T=-T(\zeta+\zeta_{g})$ and $\partial_t T_0=-T_0(\zeta_0+\zeta_{0g})$. Since the partial temperature $T_0$ depends on time through its dependence on the granular temperature, from the identity $\partial_t T_0=-T_0(\zeta_0+\zeta_{0g})$ we get the relation
\beq
\label{e1}
\left(\zeta^*+\zeta_g^*\right)\chi \tau_{\chi,0}=\tau_0 \left(\zeta_0^*+\zeta_{0g}^*-\zeta^*-\zeta_g^*\right).
\eeq
In the steady state, $\zeta_0^*+\zeta_{0g}^*=\zeta^*+\zeta_g^*=0$ and so, according to Eq.\ \eqref{e1} the derivative $\tau_{\chi,0}$ is indeterminate. As in previous works, the above problem can be fixed by using l'H\^opital's rule. Thus, we take first the derivatives with respect to $\chi$ in both sides of Eq.\ \eqref{e1} and then take the steady state condition ($\zeta_0^*+\zeta_{0g}^*=\zeta^*+\zeta_g^*=0$). After some algebra, one gets the following expression of $\tau_{\chi,0}$:
\beq
\label{e2}
\tau_{\chi,0}=\frac{A-\frac{\partial \zeta_{g}^*}{\partial \chi}}{B}=\frac{A+\frac{\mu^{3/2}\epsilon}{\chi^2}\frac{1-3x^2}{x}}{B},
\eeq
where
\beq
\label{e3}
A=-\frac{1}{2}\frac{\mu_{0g}x_0}{\chi^2 \tau_0\left[\mu_{g0}+\mu_{0g}\left(\chi \tau_0\right)^{-1}\right]}\frac{\partial \zeta_{0g}^*}{\partial x_0}=-\frac{\mu_{0g}^{3/2}\epsilon_0}{\chi^2 \tau_0^{1/2}}\frac{1-3x_0^2}{x_0},
\eeq
\beq
\label{e4}
B=-\frac{\mu^{3/2}\epsilon}{\chi\tau_0}\frac{1-3x^2}{x}-\frac{\chi}{\tau_0}A-\frac{1}{2}\frac{\zeta_{0g}^*}{\tau_0}+
\frac{m_0}{m\tau_0^2}\frac{\partial \zeta_0^*}{\partial \theta_0}.
\eeq
More explicitly,
\beq
\label{e4.1}
\chi \tau_{\chi,0}=\frac{\frac{\mu_{0g}^{3/2}\epsilon_0}{\tau_0^{1/2}}\frac{1-3x_0^2}{x_0}-\mu^{3/2}\epsilon\frac{1-3x^2}{x}}
{\frac{1-3x^2}{x}\frac{\mu^{3/2}}{\tau_0}\epsilon-\frac{\mu_{0g}^{3/2}\epsilon_0
}{ \tau_0^{3/2}}\frac{1-3x_0^2}{x_0}+\frac{1}{2}\frac{\chi \zeta_{0g}^*}{\tau_0}-
\frac{m_0\chi}{m\tau_0^2}\frac{\partial \zeta_0^*}{\partial \theta_0}}
\eeq
Once the derivative $\tau_{\chi,0}$ is known, the derivative $\tau_{\phi,0}$ can be obtained in a similar way. Its expression is  \beq
\label{e5}
\phi \tau_{\phi,0}=-\frac{\zeta_g^* \chi \tau_{\chi,0}+\tau_0\left(\zeta_g^*-\zeta_{0g}^*\right)}{\tau_0\left(\frac{\partial \zeta_{0g}^*}{\partial \tau_0}-\frac{m_0}{m \tau_0^2}\frac{\partial \zeta_0^*}{\partial \theta_0}
\right)}.
\eeq
For mechanically equivalent particles, $\mu_g=\mu_{g0}$, $\mu=\mu_{0g}$, $\epsilon=\epsilon_0$, $x=x_0$, $\tau_0=1$, and $\zeta_g^*=\zeta_{0g}^*$. Thus, according to Eqs.\ \eqref{e4.1} and \eqref{e5}, $\tau_{\chi,0}=\tau_{\phi,0}=0$ .

\section{Some technical details of the DSMC method}
\label{appF}
In this study, we investigate the numerical solution of the Boltzmann equation for a ternary mixture composed of three distinct interacting species. Our approach builds upon the framework introduced in \cite{GG22}, extending it to accommodate the dynamics of a binary granular mixture.

Particle velocities for each species $i = 1, 2, 3$ are initialized by sampling from a Maxwellian distribution at a common initial temperature $T(0)$. These velocities define the discrete representation of the velocity distribution function via a collection of $\mathcal{N}_i$ simulated (virtual) particles:

\beq
\label{APF1}
f_i^{(\mathcal{N}_i)}(\mathbf{v};t)\to \frac{n_i}{N_i}\sum_{k=1}^{\mathcal{N}_i}\delta[\mathbf{v}-\mathbf{v}_k(t)],
\eeq

where $\delta$ is the Dirac delta function and $\mathbf{v}_k$ denotes the velocity of the $k$-th particle.

Assuming a low-density regime, where collisions are treated as instantaneous binary events, the simulation can decouple free streaming and collisional stages. Given that the system remains at homogeneous states, we focus exclusively on the collisional process. The DSMC algorithm is employed as follows:

\begin{enumerate}
    \item  For each pair of species $ (i,j) $, a number of candidate collisions $ \mathcal{N}_{ij}^{\Delta t} $ is sampled over a time step $ \Delta t $. This number is given by
    \beq
    \label{APF2}
    \mathcal{N}_{ij}^{\Delta t}=\pi \mathcal{N}_i n_j \sigma_{ij}^2 g_{ij}^{\text{max}} \Delta t,
    \eeq
    where $ \sigma_{ij}=(\sigma_i+\sigma_j)/2 $ is the effective collision diameter, and $ g_{ij}^{\text{max}} $ is an upper bound for the relative velocity. A common estimate is
    \beq
    g_{ij}^{\text{max}} = C v_{ij}^{\text{th}}, \quad v_{ij}^{\text{th}} = \sqrt{\frac{2T(0)}{\overline{m}}}, \quad \overline{m} = \frac{m_i + m_j}{2},
    \eeq
    with $C$ typically set to 5 \cite[]{B94}.
    
    \item  For each pair $ (k,\ell) $, a random unit vector $ \widehat{\boldsymbol{\sigma}}_{k\ell} $ is selected uniformly over the unit sphere to define the collision axis.
    
    \item  The collision is accepted only if the relative velocity component along the selected direction exceeds a random threshold:
    \beq
    \label{APF3}
    |\widehat{{\boldsymbol {\sigma }}}_{k\ell}\cdot (\mathbf{v}_{k}-\mathbf{v}_\ell)| > \mathcal{R}(0,1)\, g_{ij}^{\text{max}},
    \eeq
    where $ \mathcal{R}(0,1) $ is a uniformly distributed random number in $[0,1]$.
    
    \item  If accepted, the post-collision velocities are computed using the inelastic scattering rules \cite{G19}:
    \begin{eqnarray}
    \label{AP24}
    \mathbf{v}_k &\to& \mathbf{v}_k - (1+\alpha_{ij})\mu_{ji}(\mathbf{g}_{k\ell} \cdot \widehat{\boldsymbol{\sigma}}_{k\ell})\widehat{\boldsymbol{\sigma}}_{k\ell}, \nonumber \\
    \mathbf{v}_\ell &\to& \mathbf{v}_\ell + (1+\alpha_{ij})\mu_{ij}(\mathbf{g}_{k\ell} \cdot \widehat{\boldsymbol{\sigma}}_{k\ell})\widehat{\boldsymbol{\sigma}}_{k\ell},
    \end{eqnarray}
    where $ \mu_{ij} = m_i/(m_i + m_j) $ and $ \alpha_{ij} $ is the restitution coefficient for collisions between species $ i $ and $ j $.

    \item The ternary nature of the system requires considering all possible species pairs $(i,j) \in \{1,2,3\} \times \{1,2,3\}$, where we label species 1 as the molecular gas, species 2 as the granular particles (grains), and species 3 as the intruder.
\end{enumerate}

    Given that both granular species (2 and 3) are present in tracer concentrations with respect to the molecular gas (1), a number of simplifications apply:
    \begin{itemize}
        \item Molecular–molecular (1–1) collisions are neglected, as the molecular gas remains in thermal equilibrium throughout the simulation and its velocity distribution is not explicitly evolved.
        
        \item Grain–molecular (2–1) and intruder–molecular (3–1) collisions are taken into account, but only the velocity of the granular particle (grain or intruder) is updated in the collision. The particles of the molecular gas are treated as part of a thermal bath.
        
        \item Grain–grain (2–2) collisions are explicitly simulated, since the concentration of grains, though low, is not negligible and can contribute to the transport properties.
        
        \item Intruder–grain (3–2) collisions are included, but only the velocity of the intruder (species 3) is updated. The grain is treated as a passive scattering particle due to the intruder’s low concentration.
        
        \item Intruder–intruder (3–3) collisions are completely neglected, given the extremely dilute concentration of intruders.
    \end{itemize}

In the present model, the molecular gas (species 1) acts as a thermal bath for the granular components (species 2 and 3), remaining in equilibrium throughout the simulation. All the relevant mechanical influence of the gas is effectively captured by the dimensionless friction (or drift) coefficient $\gamma^*$, which depends on the reduced bath temperature $T_g^*$ and the mass ratio $m/m_g$, where $m$ and $m_g$ refer to the grain and gas particle masses, respectively.

Let $N_g$ and $N$ denote the number of granular and gas particles, respectively. Since their number density ratio satisfies $N/N_g = n/n_g$, one can derive a relationship between the particle diameters $\sigma$ (for grains) and $\sigma_g$ (for gas particles). Specifically, the following constraint must hold \cite{GG22}:
\beq
\label{APF4}
\sigma_g=\left[\left(\frac{\sqrt{\pi}}{4\sqrt{2}}\frac{N}{N_g}\sqrt{\frac{m}{m_g}}\frac{1}{\phi\sqrt{T_g^*}}\right)^{1/(d-1)}-1\right]\sigma,
\eeq
which ensures consistency between the theoretical model and the simulated system. Here, $\phi$ is the solid volume fraction of the grains. Equation \eqref{APF4} follows from combining the expressions for the number densities:
\beq
\label{APF5}
n=\frac{2^{d-1}d\Gamma\left(\frac{d}{2}\right)}{\pi^{d/2}}\sigma^{-d}\phi, \quad
n_g=\frac{d \Gamma\left(\frac{d}{2}\right)}{4\pi^{(d-1)/2}}\left(\frac{m}{m_g}\right)^{1/2}\left(\frac{m}{2 T_g}\right)^{1/2}\overline{\sigma}^{1-d}\gamma,
\eeq
where $\overline{\sigma} = (\sigma + \sigma_g)/2$.

Given this dependence, once the dimension $d$, the mass ratio $m/m_g$, the reduced bath temperature $T_g^*$, and the packing fraction $\phi$ are fixed, the number of gas particles $N$ is chosen such that $N \gg N_g$ and $\sigma_g > 0$.

To simulate the Fokker--Planck model in the Brownian limiting case, each grain's velocity is updated at every time step $\delta t$ according to the Langevin-like rule
\beq
\label{APF6}
\mathbf{v} \to e^{-\gamma \delta t}\mathbf{v} + \left(\frac{6\gamma T_g \delta t}{m}\right)^{1/2} \mathbf{U}[-1,1],
\eeq
where $\mathbf{U}$ a random vector with uniform components in the interval $[-1,1]$. This update scheme reproduces the action of the Fokker--Planck operator (see Eq.\ (2.17) of the main text) in the limit $\delta t \ll \tau_{\text{coll}}$, where $\tau_{\text{coll}}$ is the mean free time between collisions \cite[]{KG14,GKG21}.

\bibliographystyle{jfm}

\end{document}